\documentclass{aa}  

\usepackage{graphicx}
\usepackage{txfonts}
\usepackage{setspace}
\usepackage{lscape}
\usepackage{caption}
\usepackage[english]{babel}
\usepackage{graphicx}
\usepackage{subfigure}
\usepackage{txfonts} 
\usepackage{textcomp}
\usepackage{natbib}
\usepackage{amssymb}
\usepackage{amsmath}
\usepackage{graphicx}
\usepackage{tabularx}
\usepackage[english]{babel}
\usepackage[T1]{fontenc}
\usepackage{multirow}
\usepackage{natbib}

\usepackage{color}

\def\HI{\ion{H}{I}}
\def\Htwo{H$_{\,2}$}
\def\OIII{[\ion{O}{III}]}

\newcommand{\Htwoi}{H$_{\,2}$ 1-0 S(1)}
\newcommand{\couno}{$^{12}$ CO (1--0)}
\newcommand{\cotre}{$^{12}$ CO (3--2)}
\newcommand{\kms}{$\,$km~s$^{-1}$}

\newcommand{\mJy}{mJy}

\newcommand{\msun}{{${\rm M}_\odot$}}
\newcommand{\msunyr}{{${\rm M}_\odot$ yr$^{-1}$}}

\newcommand{\cmsq}{cm$^{-2}$}

\newcommand{\alun}{${\rm M}_\odot\cdot$ (K km s${^{-1}}$ pc$^{-2}$)$^{-1}$}
\def\emph#1{{\sl #1}}
\newcommand{\eg}{\mbox{e.g.}}
\newcommand{\ie}{\mbox{i.e.}}

\newcommand{\pks}{\mbox PKS B1718--649}
\newcommand{\vsys}{{$v_{\rm sys}$}}

\newcommand{\co}{$^{12}$CO (2--1)}

\newcommand{\fluxco}{{$S_{\rm CO \, (2-1)}$}}
\newcommand{\mhtwo}{{$M$ (H$_{\, 2}$)}}

\newcommand{\alphac}{{$\alpha_{\rm CO}$}}

\begin{document}

   \title{ALMA observations of AGN fuelling:}
   \subtitle{ the case of \pks}

\titlerunning{ALMA observations of \pks}
\authorrunning{Maccagni et al.}

   \author{F. M. Maccagni\inst{1}$^,$\inst{2}
   \and R. Morganti\inst{1}$^,$\inst{2}
   \and T. A. Oosterloo\inst{1}$^,$\inst{2} 
   \and J. B. R. Oonk\inst{2}$^,$\inst{3}
	\and B. H. C. Emonts\inst{4}}

   \institute{ Kapteyn Astronomical Institute, University of Groningen, Postbus 800, 9700 AV Groningen, The Netherlands
   \and ASTRON, Netherlands Institute for Radio Astronomy, Postbus 2, 7990 AA, Dwingeloo, The Netherlands
   \and Leiden Observatory, Leiden University, Postbus 9513, 2300 RA Leiden, the Netherlands
   \and National Radio Astronomy Observatory, 520 Edgemont Road, Charlottesville, VA 22903, USA
                      }

 \offprints{maccagni@oa-cagliari.inaf.it}
 
  \abstract {
  	We present ALMA observations of the \co\ line of the newly born ($t_\mathrm{radio}\sim10^2$ years) active galactic nucleus (AGN), \pks. These observations reveal that the carbon monoxide in the innermost $15$ kpc of the galaxy is distributed in a complex warped disk. In the outer parts of this disk, the CO gas follows the rotation of the dust lane and of the stellar body of the galaxy hosting the radio source. In the innermost kiloparsec, the gas abruptly changes orientation and forms a circumnuclear disk ($r\lesssim700$ pc) with its major axis perpendicular to that of the outer disk. Against the compact radio emission of \pks\ ($r\sim 2$ pc), we detect an absorption line at red-shifted velocities with respect to the systemic velocity ($\Delta v = +365\pm22$\kms). This absorbing CO gas could trace molecular clouds falling onto the central super-massive black hole. A comparison with the near-infra red \Htwoi\ observations shows that the clouds must be close to the black hole ($r\lesssim 75$ pc).  The physical conditions of these clouds are different from the gas at larger radii, and are in good agreement with the predictions for the conditions of the gas when cold chaotic accretion triggers an active galactic nucleus. These observations on the centre of \pks\ provide one of the best indications that a population of cold clouds is falling towards a radio AGN, likely fuelling its activity.
}


   \maketitle
%
\section{Introduction}
\label{sec:intro_ch6}

The presence of an active galactic nucleus (AGN) in the centre of a galaxy is associated with the accretion of material onto the central super-massive black hole (SMBH), and the consequent release of energy into the host galaxy, either through radiation or relativistic jets of radio plasma, or both. The effect of the nuclear activity onto the interstellar medium (ISM) likely plays a fundamental role in regulating the star formation of the host galaxy as well as the observed relation between the masses of the bulge and the SMBH ~\citep{bower2006,croton2006,booth2009,ciotti2010,faucher2012,debuhr2012,king2015a,king2015b}. The energy released by the AGN in the ISM depends on the efficiency of the accretion onto the SMBH. However, identifying the gas that is falling onto the SMBH and characterising its kinematics and physical properties has proven to be  a particularly difficult task.

AGN are broadly classified as radio-loud when the nucleus expels a considerable fraction of the accreted energy through jets of radio plasma, otherwise they are called radio-quiet. Radio-loud AGN are divided into two categories depending on the efficiency of their accretion rate (\eg ~\citealt{best2005,best2012,alexander2012,heckman2014}). On the one hand, radiatively efficient AGN have high accretion rates (between one and ten per cent of their Eddington rate, $\lambda_\mathrm{Edd}$\footnote{$\lambda_\mathrm{Edd} =\frac{L_\mathrm{bol}}{L_\mathrm{Edd}}$. $L_\mathrm{bol}$ is the bolometric luminosity of the source and the Eddington luminosity is $L_\mathrm{Edd} = \frac{4\pi G m_\mathrm{p}c}{\sigma_T}M_\mathrm{SMBH}$; where $m_p$ is the mass of the proton, $\sigma_T$ the cross section for Thomson scattering, $G$ is the gravitational constant, $c$ the speed of light and $M_\mathrm{SMBH}$ the mass of the SMBH.}), and the SMBH expels most of the energy through radiation which ionises the surrounding ISM. On the other hand, jet-mode radio AGN have lower accretion rates ($\lambda_\mathrm{Edd}\lesssim 1\%$), and most of the energy of the AGN is deposited mechanically in the ISM by jets of radio plasma. The radio jets produce some of the most directly observable phenomena of feedback from AGN, for example inflated bubbles, cavities in the hot gas, or fast outflows of gas driven by the expansion of the radio jets (see \eg\ \citealt{fabian2012} for a review).

While radiative mode AGN are considered to be triggered mostly via mergers or secular events (\eg\ \citealt{hernquist1992,kormendy2004,comerford2009,hopkins2010,alexander2012}), it is uncertain which processes trigger jet-mode AGN. The duration of the nuclear activity ($\sim10^8$ years) is much shorter than the lifetime of the host galaxy, hence the accretion event is likely short and could occur multiple times throughout the evolution of the galaxy (\eg\ \citealt{clarke1991,jones2001,shabala2008,saikia2009,brienza2015,morganti2017}). The mechanism by which the fuelling gas is transported to the vicinity of the black hole, and the nature of the accretion flow onto the black hole must play a role  determining the efficiency of accretion onto the SMBH. 

Different accretion mechanisms have been proposed to explain the triggering of jet-mode AGN. One possibility (\eg\ \citealt{bondi1952,narayan1995,ho2009}) is that the accreting material has an inflow time much shorter than its radiative cooling time and hot gas is possibly the main fuel for the nuclear activity (\eg\ \citealt{hardcastle2007}). The simplest model for accretion of hot gas  \citep{bondi1952} assumes that the accreting material is initially distributed in a spherically symmetric geometry with negligible angular momentum. Bondi accretion does not realistically describe the circumnuclear environments of a radio AGN but it provides an estimate of the accretion rate which correlates with the power released by the radio AGN (\ie\ \citealt{allen2006,russell2013}). 

A more plausible model for the accretion mechanism of jet-mode radio AGN has been proposed based on different numerical simulations which consider that the gas cools down to temperatures lower than $10^3$ K prior to the accretion onto the SMBH. These simulations predict that the circumnuclear regions of radio AGN are rich in molecular gas distributed in a clumpy disk, with kinematics characterised both by rotation and turbulence (\eg~\citealt{wada2005,wada2009}). Accretion of small ($30\leq r \leq 150$ pc) clouds of gas is a process that could explain both the low accretion rates typical of these active nuclei as well as the short time-scale of the fuelling of the radio AGN ~\citep{nayakshin2012,gaspari2013,hillel2013,king2015b,gaspari2015b,gaspari2017a,gaspari2017}. In this process, dense clouds of cold gas form within a hot medium because of turbulence, and only within a few Bondi radii the chaotic inelastic collisions between the clouds are strong enough to cancel their angular momentum and the clouds accrete onto the SMBH. 

Over the past few years, different observations have allowed us to characterise the kinematics and the physical conditions of the cold atomic and molecular gas in the circumnuclear regions of a handful of radio AGN (\eg\ \citealt{davies2009,davies2011,muller2013,combes2013,combes2014,garciaburillo2014,viti2014,davies2014,morganti2015,martin2015,dasyra2016,garciaburillo2016,tremblay2016,russell2016,espada2017}). This enabled us to study the effect of the active nucleus on the surrounding molecular medium and, to some degree, to trace gas that may contribute to fuel the central SMBH. Hints that chaotic cold accretion can fuel active galactic nuclei have been found in different objects. For example, \mbox{PKS 2322-12}~\citep{tremblay2016}, \mbox{NGC 5044} \citep{david2014}, \mbox{PKS 0745-191} \citep{russell2016} show molecular clouds falling towards the active nucleus, that, if located close to the SMBH, could possibly accrete onto it. Clouds of \HI\ gas detected through absorption could contribute to sustain the nuclear activity of, for example, \mbox{NGC 315} \citep{morganti2009}. Observations of the molecular gas in a few young radio sources (\eg~\pks, \citealt{maccagni2016}; \mbox{4C +31.04}, \citealt{garciaburillo2007};~\mbox{3C 293}, \citealt{labiano2013}; Centaurus A, \citealt{espada2017}) infer a clumpy circumnuclear disk and find indications that the cold gas may contribute to fuel the central radio source. 

In most AGN where indications of accretion of cold gas have been found, it has been very difficult to determine the distance of the in-falling material from the SMBH, and to understand under which physical conditions of the ISM a radio AGN can be triggered. Often, the identification of the in-falling material is subject to many uncertainties, and the radio jets have already expanded for hundreds of parsecs into the ISM, possibly changing the physical conditions of the gas compared to before the nuclear activity was triggered. To study the fuelling of the central nucleus, it is important to select a radio AGN where the activity did not have time to perturb the surrounding ISM. The ideal candidate would be a very young radio source, where the radio jets extend for at most a few parsecs, and their impact on the ISM is limited.  

\begin{figure}
	\begin{center}
		\includegraphics[clip,width=\columnwidth]{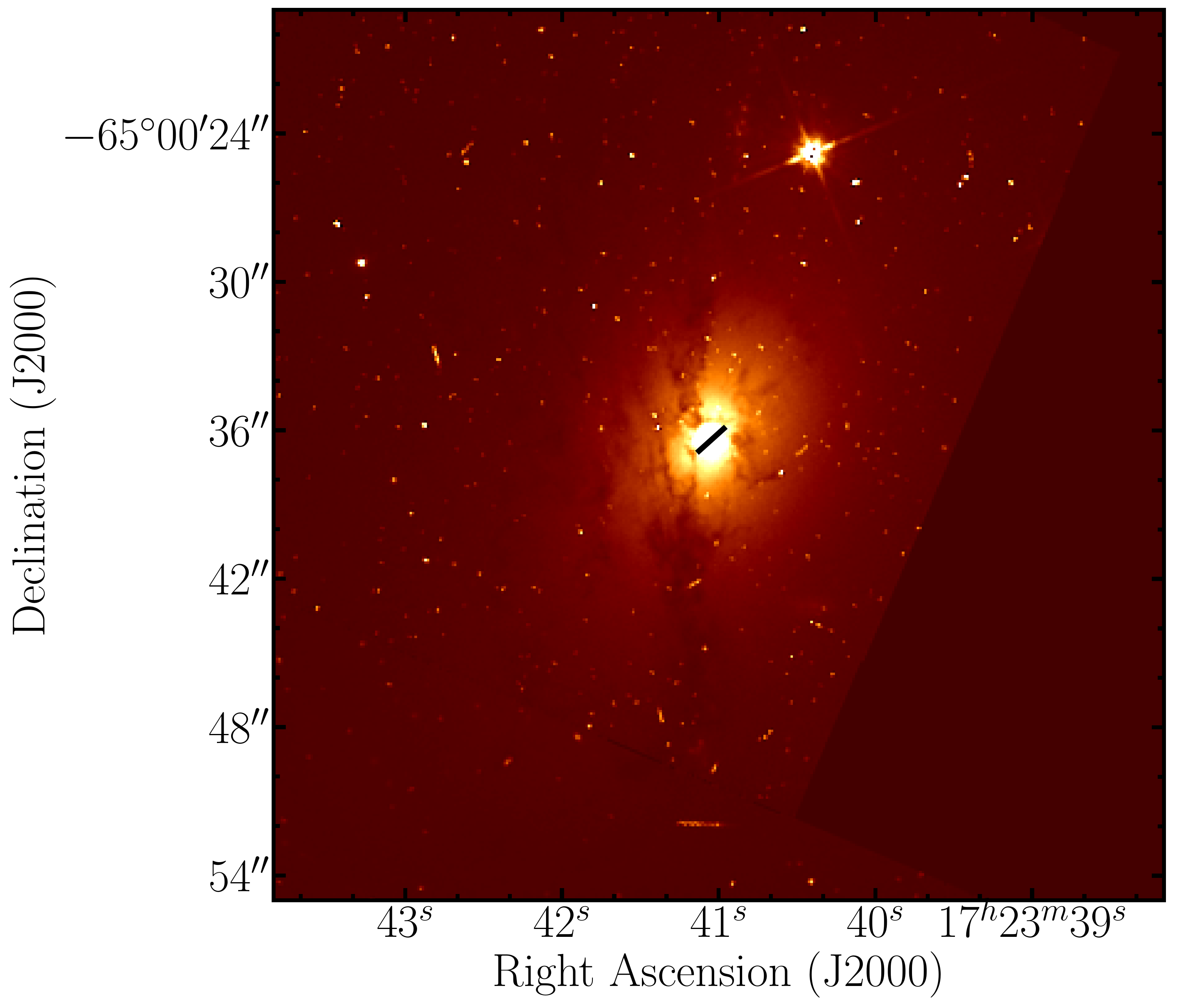}
		\caption{Hubble Space Telescope WFPC2 image of \pks\ within the field of view of the ALMA telescope. The black line shows the position angle of the radio AGN ($PA_\mathrm{radio} = 135^\circ$).}
		\label{fig:hst}
	\end{center}
\end{figure}

Young AGN with compact radio jets (\ie\ the jets are embedded within the host galaxy) are rich in atomic and molecular gas compared to other more evolved AGN (\eg~\citealt{emonts2010,guillard2012,gereb2015,maccagni2017}). The kinematics of the gas in the proximity of the AGN seems to be more unsettled than in older AGN and likely fuelling is still on-going (\eg~\citealt{allison2012,gereb2015,maccagni2017}). These observations lack the spatial resolution to resolve individual cloud-like structures, but their results strongly suggest that compact and young radio AGN are the best candidates to study the triggering and fuelling of low-efficiency active nuclei. 

In the following section, we introduce the main characteristics of \pks, which is a nearby (z=$0.01442$)\footnote{Throughout this paper we use a $\Lambda$CDM cosmology, with Hubble constant $H_0$ = 70\kms\ Mpc$^{-1}$ and $\Omega_\Lambda = 0.7$ and $\Omega_{\rm M} = 0.3$. At the distance of \pks\ (62.4 Mpc): $1\arcsec = 294$ pc.} young radio AGN, thus, an ideal candidate to study the triggering of young active galactic nuclei.

\subsection{A baby radio source: \pks}

\pks\ is one of the youngest sources ($t_{\rm{radio}}\sim10^2$ years, \citealt{giroletti2009}) among nearby radio AGN. The radio source is compact, extending  only 2 pc and it is classified as a Giga-Hertz Peaked Spectrum source (GPS,~\citealp{tingay1997,tingay2002}). Previously, we observed the source in atomic hydrogen (\HI) with the Australia Telescope Compact Array (ATCA) and in warm molecular hydrogen (\Htwo, $2.12 \mu$m) with SINFONI. The results are shown in \citet{maccagni2014,maccagni2016}. 

\pks\ is hosted by an early-type galaxy and is embedded in a large-scale \HI\ disk with regular kinematics~(\citealt{veron1995,maccagni2014}). The regularity of the \HI\ disk rules out interaction events as responsible for the fuelling of the radio source~\citep{maccagni2014}.

Figure~\ref{fig:hst} shows the optical emission of  the central regions of \pks\ observed by the Hubble Space Telescope WFPC2. A dust-lane oriented north-south crosses the central bulge. Clouds of dust are detected also in other regions of the field. The black line in the figure shows the orientation of the compact radio emission of the AGN (PA$_\mathrm{radio} = 135^\circ$, $r_{\mathrm radio}\sim2$ pc, ~\citealt{tingay1997}).

The observations of the \Htwoi\ line emission at $2.12\,\mu$m \citep{maccagni2016} show that in the innermost kilo-parsec warm \Htwo\ is rotating in a disk that abruptly changes its orientation at $r\sim650$ pc. At outer radii, the disk is oriented north-south and it follows the rotation of the other components of the galaxy (e.g. stars, \HI\ disk), while at radii $\lesssim650$ pc the disk is oriented along the east-west direction. The inner circumnuclear disk is clumpy and is characterised by rotation around the centre.

However, in the innermost regions of this disk, both the \HI\ and the \Htwo\ gas show a component with complex kinematics deviating from the regular rotation of the disk. Against the compact radio continuum emission, we detect two \HI\ absorption lines with opposite velocities with respect to the systemic velocity of the galaxy. These lines trace two clouds of cold gas which are not rotating within the galactic disk. The clouds are small (M$_{\mathrm H_{\mathrm I}}$ $< 10^7$ M$_\odot$), otherwise they would have been detected in emission. As discussed in \citet{maccagni2014}, these clouds may belong to a larger population of clouds surrounding the radio source and may contribute to its fuelling. In the innermost hundreds of parsecs, a component of \Htwo\ is also found to deviate from the regular rotation of the disk with red-shifted velocities with respect to the systemic ($\gtrsim+150$\kms, \citealt{maccagni2016}). Given the clumpiness of the ISM surrounding the radio source, it is likely that the redshifted \Htwo\ and the absorbed \HI\ gas belong to the same nuclear components, in front of the radio source within the innermost few hundred parsec. If this were the case, since the velocities of the \Htwo\ gas are red-shifted, we would observe molecular hydrogen directly falling towards the SMBH. 

The spatial resolution of the $2.12\,\mu$m observations does not allow us to further constrain the kinematics of the molecular hydrogen close to the SMBH, neither to understand if and how molecular clouds could fuel the newly born radio AGN. Nevertheless, complementary observations also suggest that gas in the form of dense clouds may actively contribute to the accretion onto the SMBH. In the optical band, \pks\ is a weak AGN, classified as a LINER. From the study of the optical emission lines, \cite{filippenko1985} suggested that the circumnuclear environment of PKS is populated by dense clouds ($\sim 1.5 \times 10^{21}$ cm$^{-2}$). A clumpy circumnuclear ISM could also explain the variability of the peak of the radio spectrum due to free-free absorption~\citep{tingay2015}. While relatively weak in the optical, \pks\ has an excess of X-ray emission. This could be due to an extended thermal component typical of complex gaseous environments \citep{siemiginowska2016}. Moreover, \pks\ was the first compact young radio AGN where $\gamma$-ray emission has been detected~\citep{migliori2016}. 

In \pks\ we may be witnessing the signatures of clouds falling towards the AGN, actively contributing to its feeding, through the \HI\ lines and the broad red-shifted molecular component \citep{maccagni2014,maccagni2016}. If this was the case, higher resolution and sensitivity observations of the molecular gas in the central regions of the galaxy could provide us a detailed view of the interplay between the gas and the nuclear activity, and to understand how this radio AGN was triggered. 

In this paper, we present results from the ALMA observations of the \co\ line of the central $15$ kpc of \pks. In Sect.~\ref{sec:obs_ch6}, we describe the observations and the data reduction. In Sect~\ref{sec:res_ch6}, we reveal the complexity of the distribution and kinematics of the molecular gas. In Sect.~\ref{sec:indisk}, we focus on the study of kinematics of the gas in the circumnuclear ($r<700$ pc) regions. In Sect.~\ref{sec:red_abs}, we analyze the kinematics of the \co\ gas detected in absorption against the compact radio AGN. In Sect.~\ref{sec:masses}, we determine the masses of the molecular gas in different regions of \pks. In Sect.~\ref{sec:h2}, we compare the distribution and kinematics of the \co\ with the warm \Htwo\ \citep{maccagni2016}. In Sect.~\ref{sec:infall}, we discuss evidence for infalling clouds towards the radio source, and in Sect~\ref{sec:accretion} we analyze how accretion of these clouds onto the SMBH could sustain the radio nuclear activity of \pks. Section~\ref{sec:conc_ch6} summarizes our results and conclusions.

\section{Observations}
\label{sec:obs_ch6}
The \co\ line of \pks\ was observed by ALMA during Cycle 3 in September 2016 (ID: 2015.1.01359.S, PI: F. M. Maccagni). The observing properties are summarized in Table~\ref{table:ALMA}. The observations cover the innermost 15 kpc of \pks\ with a single-pointing field of view (FoV) of $\sim 51.2\arcsec$. The ALMA extended antenna configuration ($C36-6$, maximum baseline of $16$ km) allows us to reach maximum spatial resolution of $0.22\arcsec$, and is sensitive to a maximum recoverable scale of $2\arcsec$. At the redshift of \pks, the \co\ line is found at a frequency of $\nu_{\rm obs} = 227.591$ GHz. The Band 6 receivers were used for the observations. The spectral window of the observation was $1.875$ GHz, divided in $240$ channels of $7.81$ MHz ($\sim10.3$\kms). The duration of the observation, including the time for calibration, was 3.4 hours. The initial calibration was done in {\tt CASA}~\citep{mcmullin2007} using the ALMA data reduction scripts. These calibrated {\em uv}-data were subsequently exported to {\tt MIRIAD}~\citep{sault1995} which was used to perform additional self-calibration of the continuum point source. This improved the quality of the images and data cubes. All further reduction steps (continuum subtraction, mapping/cleaning) were also done in {\tt MIRIAD}.

\begin{table}[tbh]
	\caption{Specifications of the ALMA final data cube}             
	\label{table:ALMA}      
	\centering                          
	\begin{tabularx}{\columnwidth}{X c }        
		\hline\hline                 
		Parameter & Value \\    
		\hline                        
		Field of view & $51.2\arcsec \times 51.2\arcsec$ ($15\times 15$ kpc) \\ 
		Synthesized beam & $0.28\arcsec \times 0.19\arcsec$ ($82\times 82$ pc)\\
		Velocity Resolution  &  48~\kms \\
		r.m.s noise per channel & 0.09~\mJy\ beam$^{-1}$  \\
		\hline                                   
	\end{tabularx}
\end{table}

The data cubes were made using various Briggs weightings~\citep{briggs1999} to explore which one would allow us to better image the molecular gas. In this paper, we show the data cube obtained using natural weighting and smoothed to a velocity resolution of $48$\kms, except when indicated otherwise. The r.m.s. noise per channel of the data cube is $0.09$~\mJy\ beam$^{-1}$. The size of the synthesized beam of the final data cube is $0.28\arcsec \times 0.19\arcsec$ (PA $= 41^\circ$), which corresponds to a spatial resolution of $\sim 82$ pc.

A continuum image was also produced from the {\em uv}-data using uniform weighting. The r.m.s noise of the continuum image is $0.1$ \mJy\ beam$^{-1}$ and the restoring beam is $0.24\arcsec \times 0.15\arcsec$ (PA $= 37^\circ$). Since the emission of the AGN is strong at $230$ GHz ($S_\mathrm{cont}=303$ mJy), the quality of the continuum image is limited by the dynamic range. The continuum source is unresolved, as expected from its measured linear size at cm wavelengths ($r_\mathrm{radio}\sim 2$ pc). The $230$ GHz continuum flux found in the ALMA observations is about a factor two lower than what expected from the extrapolation of the data at frequencies of a few GHz (\eg\ \citealt{tingay1997,tingay2002,giroletti2009}). Likely, this occurs because the radio source is known to be variable \citep{tingay2015} and  monitoring observations (Moss et al. in prep) show that on the time scale of a year, the flux of the source varies by more than 50\%.

\section{Results}
\label{sec:res_ch6}
In this section, we present the analysis of the data cube of the observations of the \co\ line of \pks. The main results are that the gas detected in emission reveals a complex clumpy distribution overall dominated by rotation (Sect.~\ref{sec:disk}) and that against the central compact radio continuum, \co\ gas is detected in absorption at red-shifted velocities ($\sim+365$\kms) with respect to the systemic velocity (Sect.~\ref{sec:red_abs}).

\begin{figure}[tbh]
	\begin{center}
		\includegraphics[trim = 0 0 0 0, clip, width=\columnwidth]{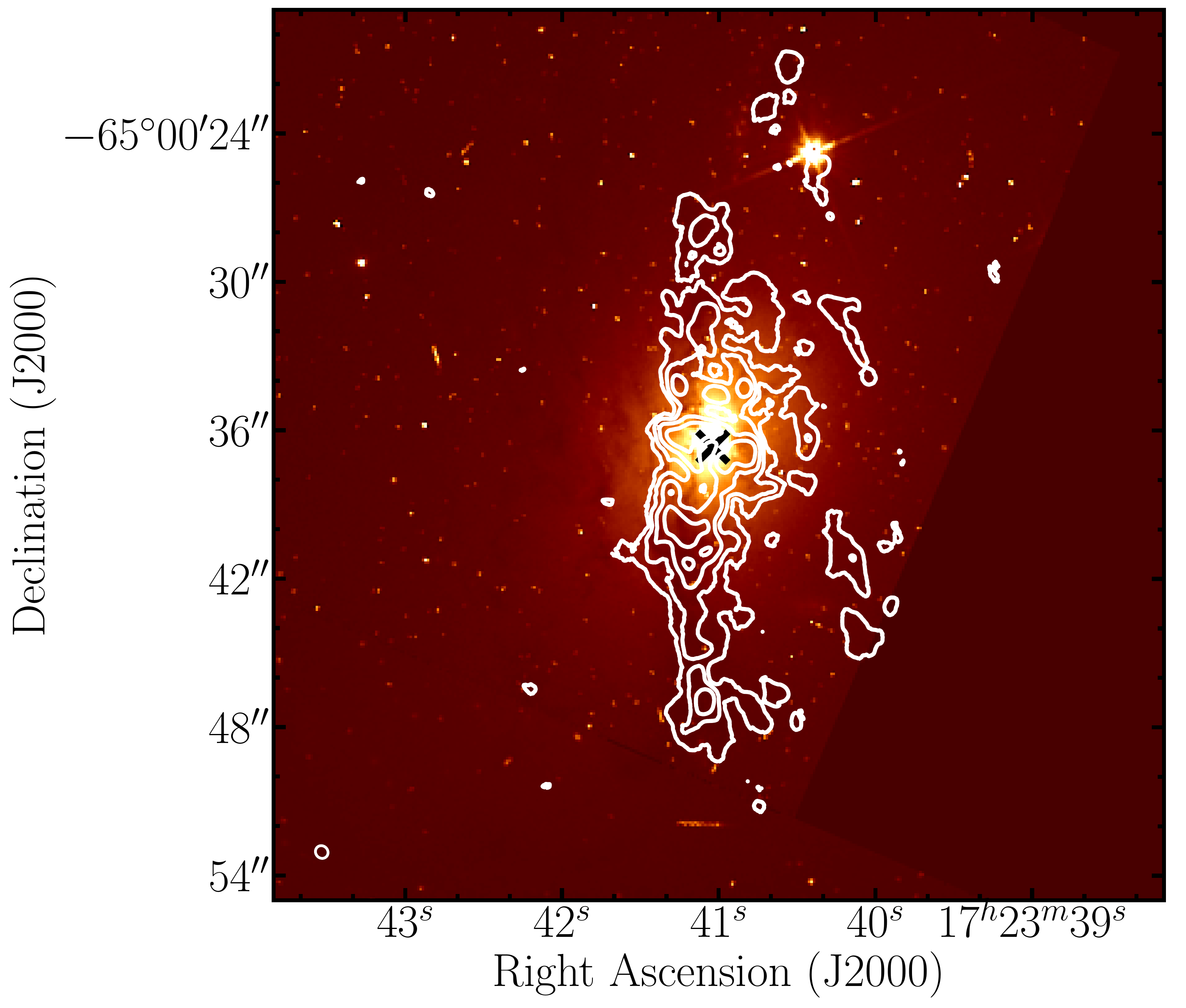}
		\caption{Hubble Space Telescope WFPC2 image of \pks\ with overlaid the contours of the total intensity map of the \co\ line detected in the innermost $15$ kpc. Contour levels are $3,6,12$ and $18\sigma$. This and all intensity maps shown in this paper are not corrected for the primary beam of ALMA. The $230$ GHz continuum emission of the radio AGN is indicated with a black cross.  The resolution of the total intensity map is $0.5\arcsec$. The beam is shown in white in the bottom left corner.}
		\label{fig:mom0}
	\end{center}
\end{figure}

\subsection{Molecular clouds rotating in a warped disk}
\label{sec:disk}

The ALMA observations of \pks\ show that the innermost $15$ kpc are rich in molecular gas distributed in clouds and filamentary structures. Figure~\ref{fig:mom0} displays the total intensity map of the \co\ line emission (extracted from a data cube with $0.5\arcsec$ of spatial resolution) overlaid on the optical image taken with the Hubble Space Telescope WFPC2 camera. Overall, the \co\ emission is located along the the dust lane visible in the optical band and oriented north-south, and in the circumnuclear regions, close to the radio AGN. The distribution of the carbon monoxide is clumpy, with many molecular clouds spatially resolved in the field of view. From the total intensity map, corrected for the primary beam of ALMA, we measure the total flux of the \co\ line, $S_{\rm CO} \Delta v = 174\pm17.4$ ~Jy~km~s$^{-1}$. We estimate that the error in the flux is $\sim 10\%$, to allow for uncertainties in the flux of the gain calibrator.

\begin{figure*}[tbh]
	\begin{center}
		\includegraphics[trim = 0 0 0 0, clip,width=0.44\textwidth]{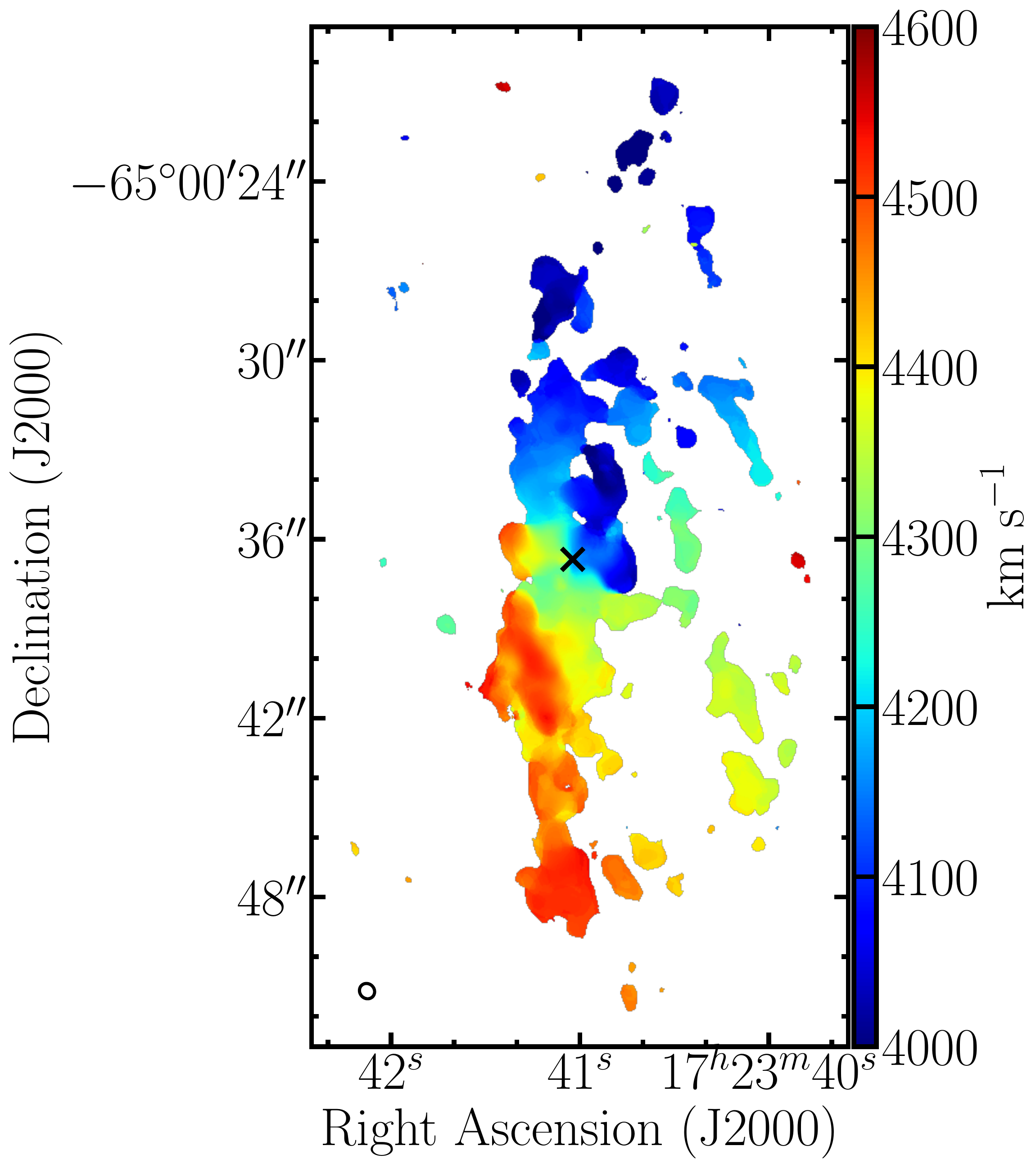}
		\includegraphics[trim = 0 0 0 0, clip,width=0.48\textwidth]{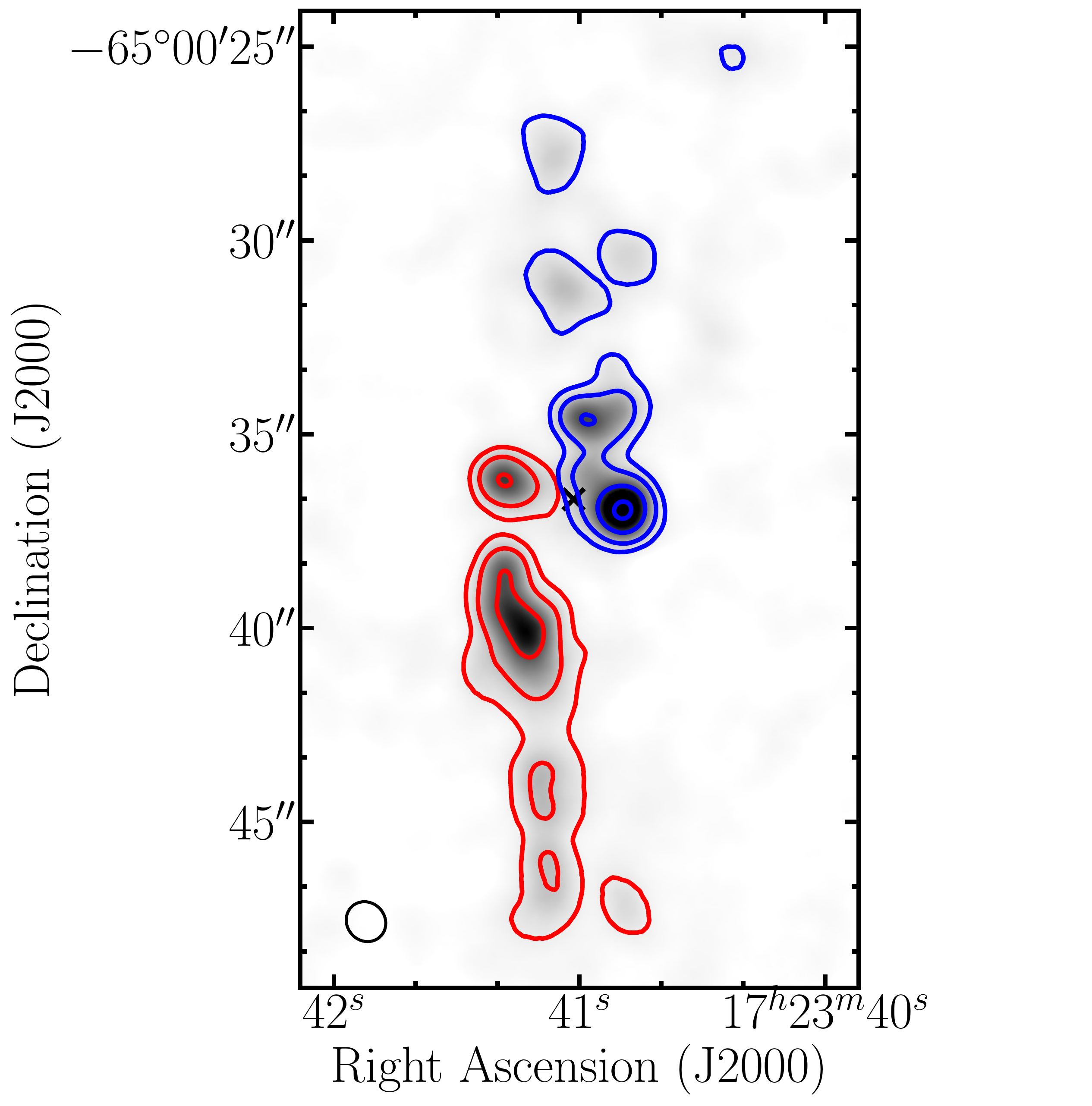}
		\caption{{\em Left Panel}: Velocity field of the \co\ line detected in the innermost $15$ kpc of \pks. The systemic velocity ($\sim 4250\pm 20$\kms) has green colours. The radio continuum emission is indicated by a black cross. The beam is shown in black in the bottom left corner. {\em Right Panel}: Superposed channel maps for the \co\ gas at velocities displaced roughly $190$\kms on either side of the systemic velocity. Upper contours, in blue, show gas centred at $4086$\kms while lower contours, in red, show gas centred at $4444$\kms.}
		\label{fig:mom1}
	\end{center}
\end{figure*}

Figure~\ref{fig:mom1} (left panel) shows the velocity field of the \co\ line emission extracted from the same data cube as the total intensity map. Overall, the kinematics of the gas is dominated by rotation. The observed velocity ranges between $\sim 3950$\kms\ and $\sim 4600$\kms\ along the north-south axis. We measure the systemic velocity of the \co\ as the mean point of this range of velocities, $4250\pm20$\kms. This velocity is comparable to the systemic velocity measured from the kinematics of the \HI\ gas (\vsys$= 4274\pm7$\kms). The position angle of the disk and the direction of rotation is the same as the \HI\ at outer radii (between $3$ and $8$ kpc) suggesting that overall the \co\ is following the rotation of the other components of the galaxy. 

The observations of the atomic and molecular hydrogen suggested that the cold gas of \pks\ is distributed a warped structure. Here, we analyse a number of different signatures indicating that this is the case. The left panel of Fig~\ref{fig:mom1} shows the velocity field of the \co\ gas seen by ALMA. Moving towards the centre, the kinematic major axis changes orientation from north-south in the outer regions to almost east-west in the inner part of the gas distribution. The figure suggests that the carbon monoxide in the circumnuclear regions and in the dust lane are connected in velocity and are distributed in a warped disk. The right panel of Fig.~\ref{fig:mom1} shows in red contours the \co\ emission detected at $4440$\kms\ and in blue contours the emission detected at $4100$\kms, extracted from a data-cube at the spatial resolution of 1\arcsec. These two channels have opposite velocities with respect to the systemic velocity of the galaxy and show a good symmetry in the distribution of the \co\ emission. This panel shows that the velocity gradient in the central regions is perpendicular to that of the larger disk.  This feature is very similar to that found in the prototypical warped galaxy \mbox{NGC 3718}~\citep{sparke2009}, even though over larger spatial scales. In this galaxy, \cite{sparke2009} showed that the \HI\ gas is distributed in a single extremely warped disk, which varies its position angle while its inclination remains fairly edge-on. Likely, also the molecular gas of \pks\ is distributed in a similar warped disk.

\begin{figure}[tbh]
	\begin{center}
		\includegraphics[clip,width=\columnwidth]{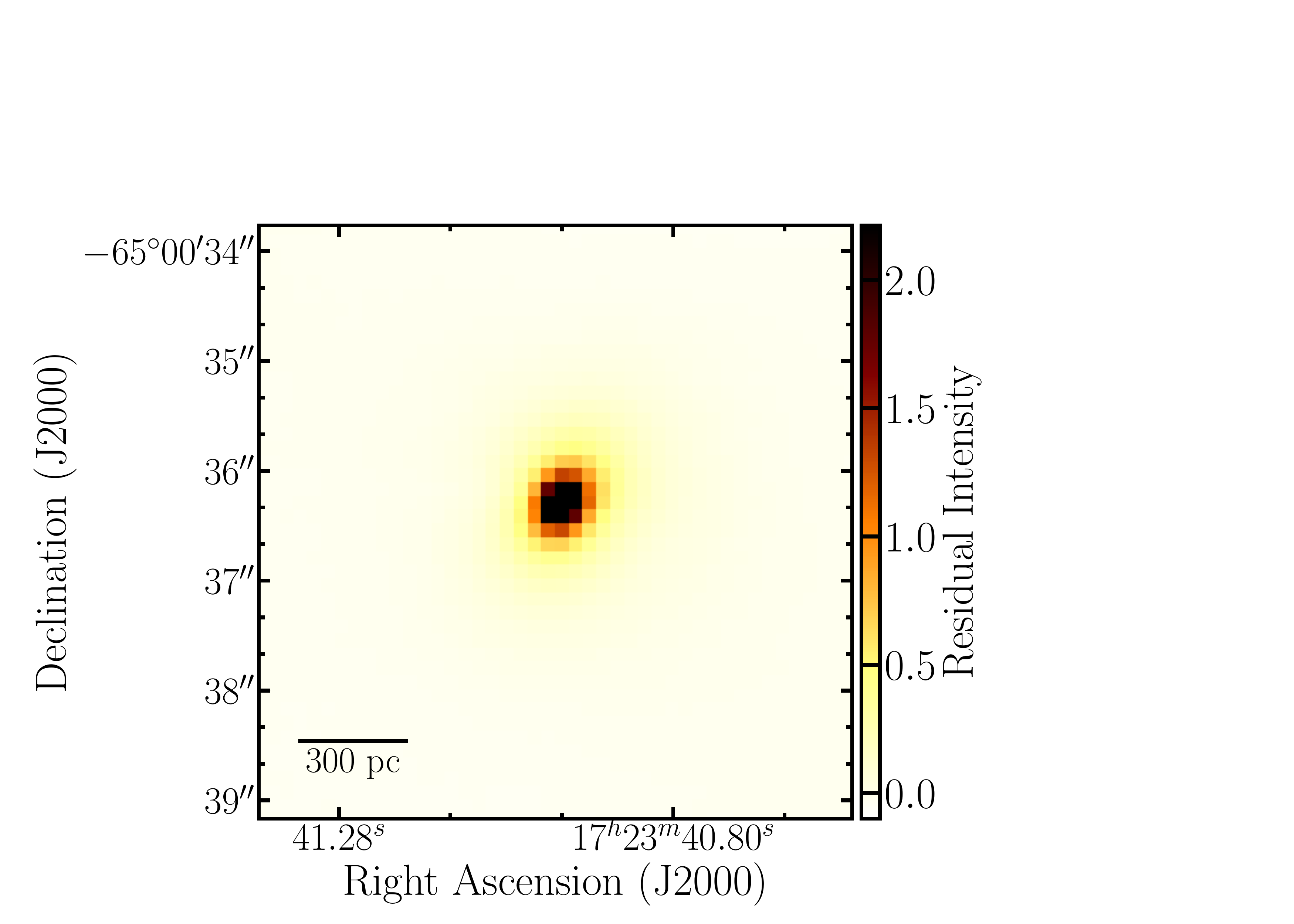}
		\caption{Map of the residuals between the light distribution of the stellar body, seen by SINFONI~\citep{maccagni2016} and a model with isophotes of constant ellipticity ($\epsilon = 1-b/a = 0.18$) and position angle (PA$=150^\circ$). No indications of the presence of a bar are found, the only significant deviations from an elliptical light distribution are found in the innermost $300$ pc, where a spherical component may be present.}
		\label{fig:stars}
	\end{center}
\end{figure}

Even though the kinematics of the molecular gas of \pks\ are consistent with a warped disk, bars in the stellar distribution can induce non-circular motions which can mimic the kinematics of a warped disk. We investigate whether a bar could be present, studying the light distribution seen in our SINFONI data \citep{maccagni2016}. At radii larger than a few kilo-parsecs, the light distribution is well modelled using elliptical isophotes (ellipticity $\epsilon = 1-b/a = 0.18$ and position angle, $PA=150^\circ$) and a deVaucouleur's law \citep{veron1995}. We subtract this model from the light distribution seen the innermost regions of \pks. The residuals (Fig.~\ref{fig:stars}) show that no signature of a bar is found. The figure also shows that the only deviations from the model are found in the centre ($r\lesssim300$ pc), and suggests that a small spherical component may be present there.

\subsection{A clumpy circumnuclear disk}
\label{sec:indisk}

\begin{figure}[tbh]
	\begin{center}
		\includegraphics[clip,width=\columnwidth]{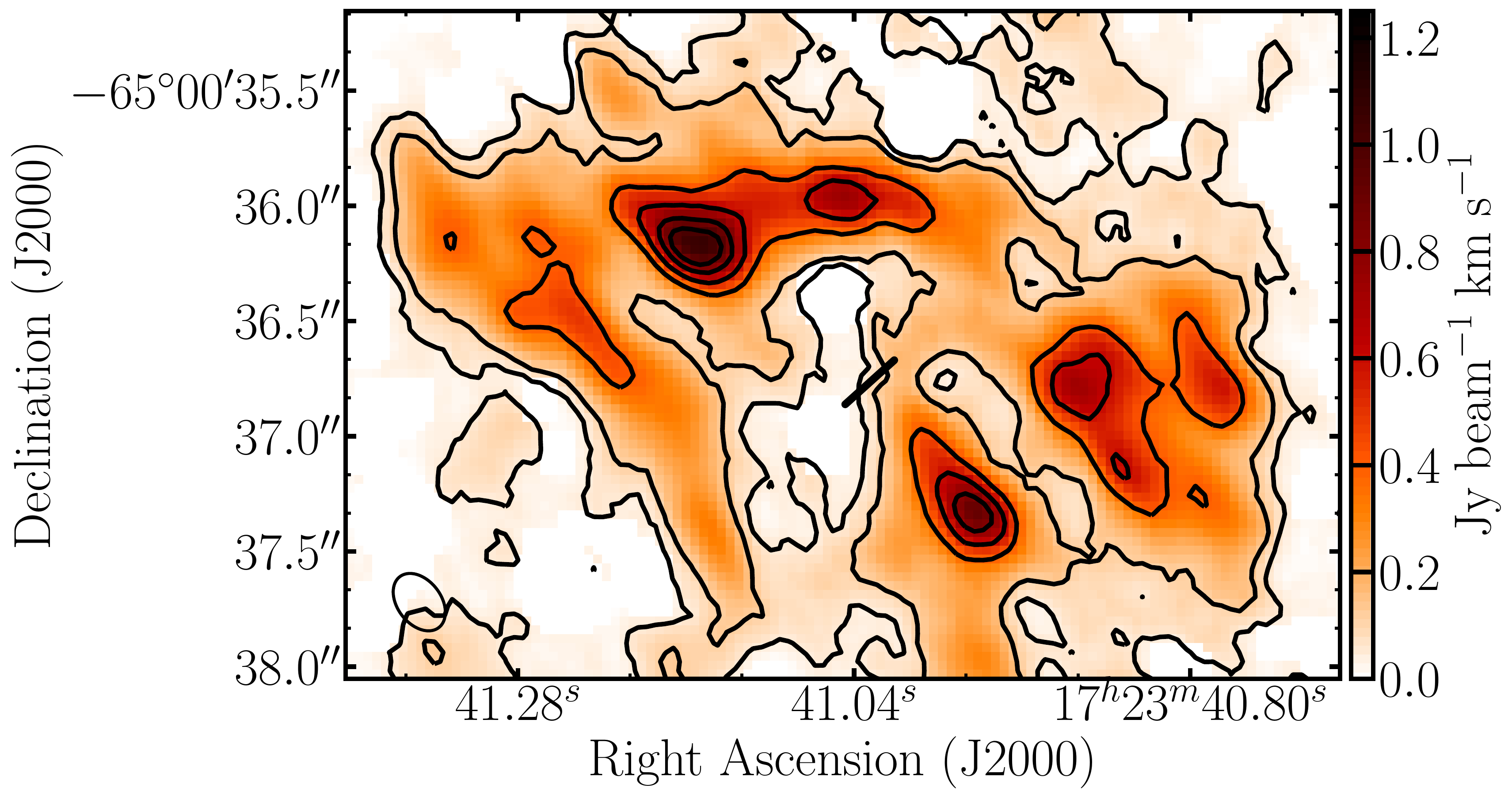}
		\caption{Total intensity map of the \co\ line detected in the innermost $1.5$ kpc of \pks. Contour levels are $3,5,8,12,15$ and $18\sigma$ ($1\sigma = 0.14$~Jy beam$^{-1}$\kms). The black line shows the position angle of the radio AGN. The beam ($0.2$\arcsec) is shown in black in the bottom left corner.}
		\label{fig:indisk_mom0}
	\end{center}
\end{figure}

Figure~\ref{fig:indisk_mom0} shows the total intensity map of the \co\ emission line in the circumnuclear regions of \pks. The rotating structure is oriented along the east-west direction (PA $=72^\circ$), it has a radius of $700$ pc. The radio source is almost orthogonal to the position angle of the major axis of the disk, $PA=135^\circ$~\citep{tingay1997,tingay2002}. The circumnuclear disk shows a clumpy structure embedded in a more diffuse medium. Within the disk, we resolve spatially and in velocity a few clouds that have approximate sizes of $r\lesssim 120$ pc. In the very centre, where the radio AGN is located (black cross in the figure), we do not detect \co\ in emission at the sensitivity of our observations. This implies a $3\sigma$ upper limit to the column density of the the molecular hydrogen of $N_{\mathrm H_2}\sim4.8\times10^{21}$\cmsq, for a Galactic conversion factor (see Sect.~\ref{sec:masses}).

The total flux density (corrected for the primary beam of ALMA) of the circumnuclear disk is $S_{\rm CO} \Delta v =29.5\pm 2.95$~Jy km s$^{-1}$ (see Table~\ref{table:em}).

\begin{figure}[tbh]
	\begin{center}
		\includegraphics[clip,width=\columnwidth]{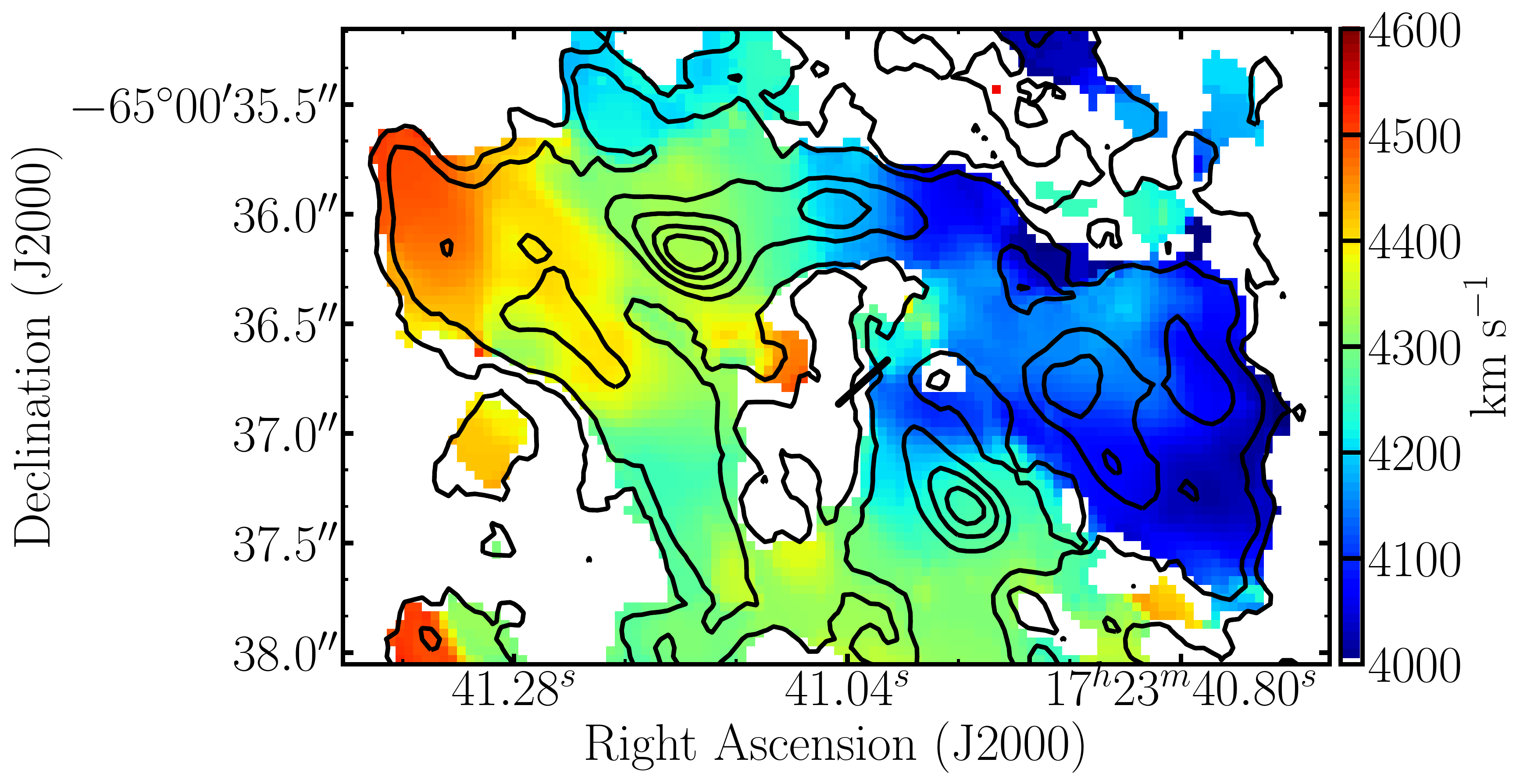}
		\caption{Velocity field of the \co\ line detected in the circumnuclear regions of \pks. Green velocities mark approximately the systemic velocity of the galaxy ($v_{\rm sys}=4250\pm20$\kms). Contour levels are as in Fig.~\ref{fig:indisk_mom0}. The black line shows the position angle of the radio AGN. The beam ($0.2$\arcsec) is shown in black in the bottom left corner.}
		\label{fig:indisk_mom1}
	\end{center}
\end{figure}

Figure~\ref{fig:indisk_mom1} shows that the circumnuclear disk is dominated by rotation. The velocity field, as well as the single channel maps, suggest the disk is approximately edge-on ($i\sim 60^\circ-80^\circ$).

Figure~\ref{fig:pv_spec} shows a position-velocity diagram taken along the major axis of the circumnuclear disk. The smooth velocity gradient along the x-axis clearly shows the disk is overall regularly rotating and has a radius of $700$ pc. In the innermost $\sim 300$ pc, the gas in the disk spans a broader range of velocities than at larger radii. This may suggest an increase in the velocity dispersion of the gas in the central regions, even if part of this broadening of the velocity gradient can be explained by the geometry of the system, because the disk is observed edge-on.

\subsection{The \co\ detected in absorption}
\label{sec:red_abs}
\begin{figure}
	\begin{center}
		\includegraphics[trim = 0 0 0 0, clip,width=\columnwidth]{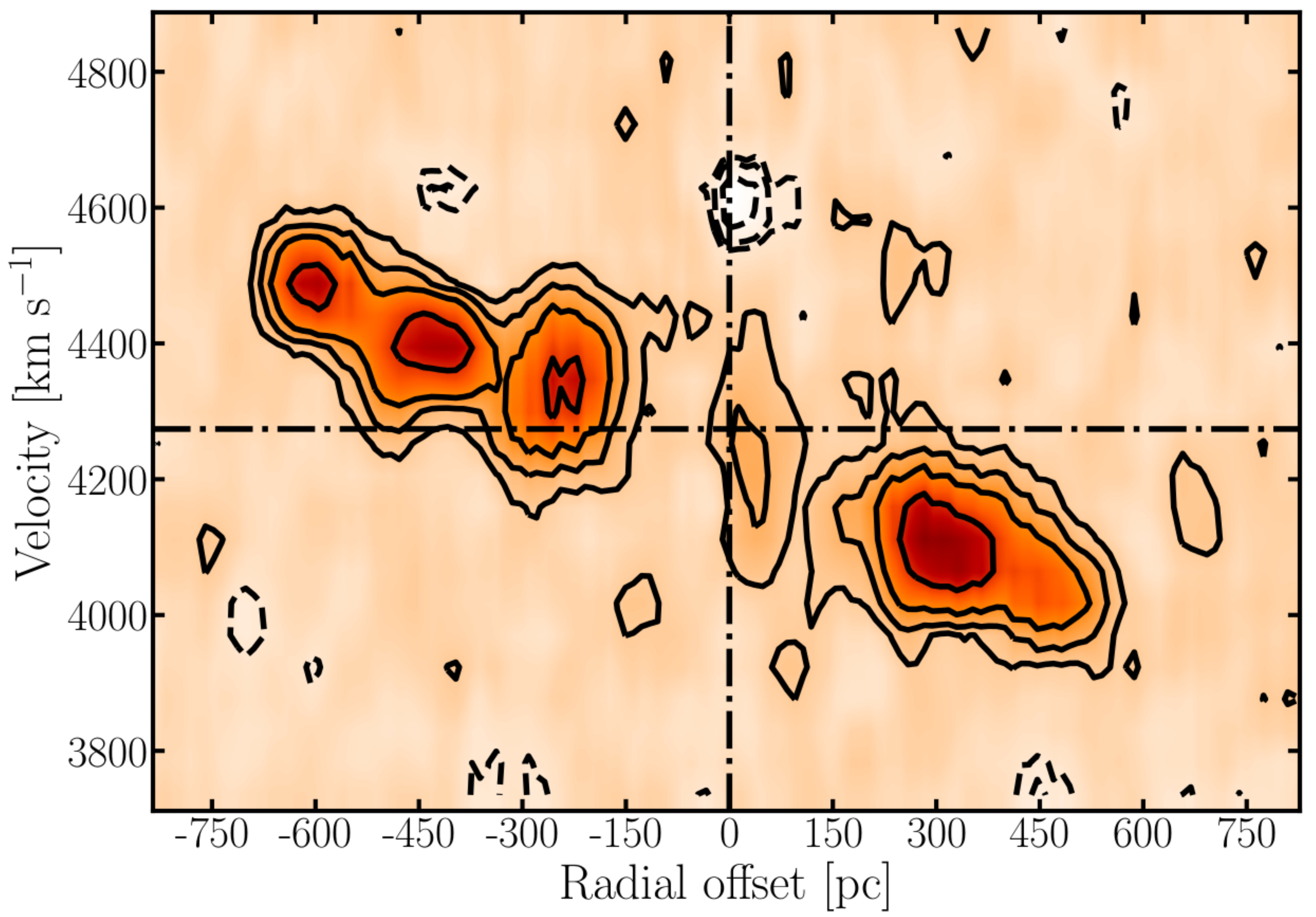}
		\caption{Position-velocity diagram taken along the major axis of the circumnuclear disk of \pks\ (PA $=72^\circ$), with a slit of $0.2$\arcsec. The dashed horizontal line shows the systemic velocity of the source. The vertical dashed line shows the position of the radio continuum sources. Contour levels are $-5,-3,\,\,-2\sigma$, in white, and $3,6,12,18,24\sigma$, in black. Absorption is detected with $\sim7\sigma$ significance at $v\sim 4615$\kms and zero offset.}
		\label{fig:pv_spec}
	\end{center}
\end{figure}

\begin{figure}
	\begin{center}
		\includegraphics[trim = 0 0 0 0, clip,width=\columnwidth]{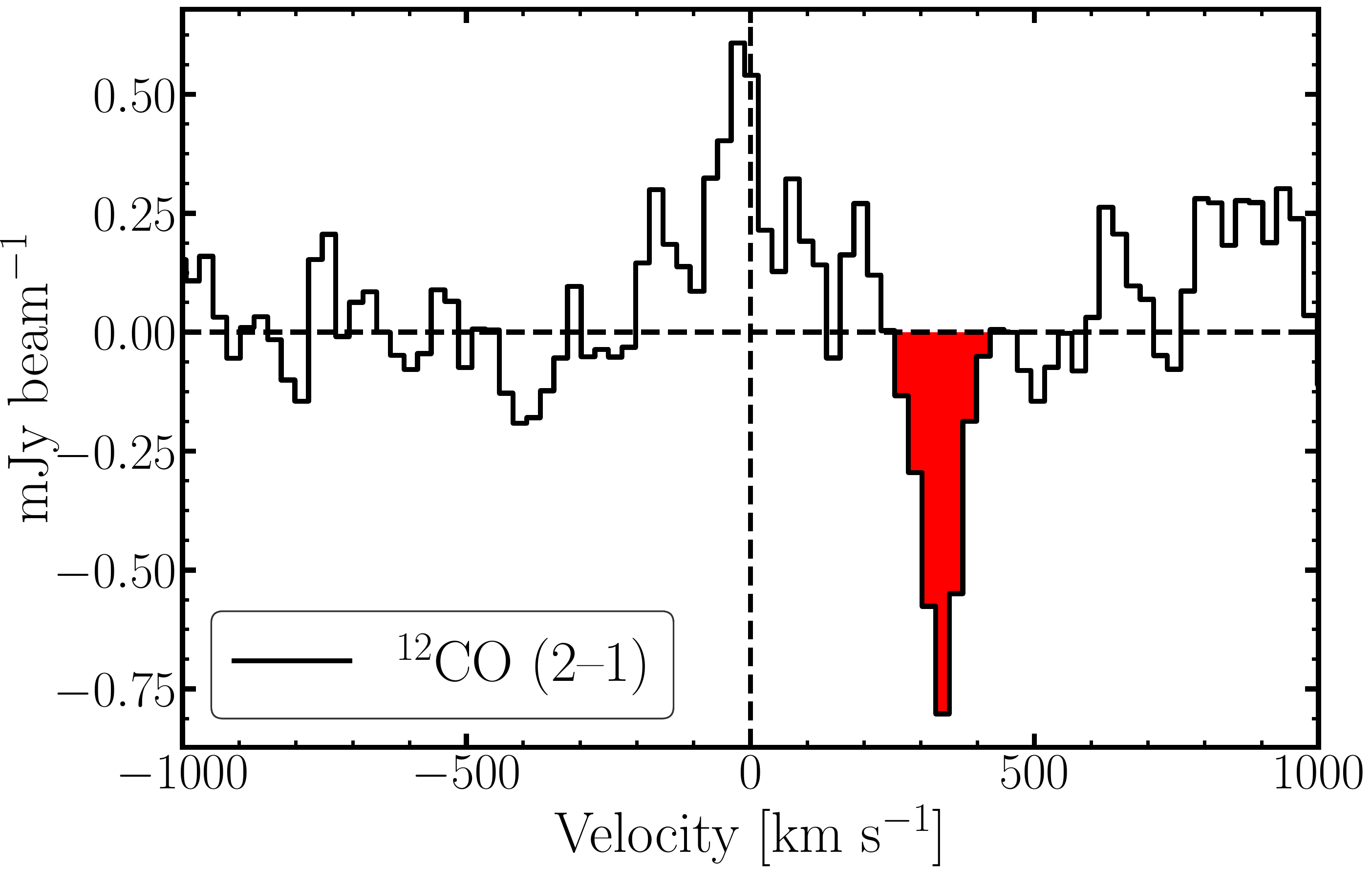}
		\caption{Spectrum extracted against the radio continuum emission of \pks. \co\ is detected in absorption at red-shifted velocities with respect to the systemic ($v=+365\pm20$\kms). The peak of the line is detected with $\sim 7\sigma$ significance. The red shaded regions shows the integrated flux of the line.}
		\label{fig:abs_line}
	\end{center}
\end{figure}

Figure~\ref{fig:pv_spec} shows that the \co\ is detected in absorption against the radio AGN, at $v=+4615\pm20$~\kms (in black dashed contours).  To investigate the line profile in more detail, in Fig.~\ref{fig:abs_line} we show the spectrum  extracted against the radio continuum emission from a data cube at a higher spectral resolution ($20$\kms). The noise of the spectrum is $1.12$~mJy~beam$^{-1}$. The peak of the absorption line (S$_\mathrm{peak}=-0.813$ mJy) is detected with $\sim 7\sigma$ significance. The total absorbed flux is $2.59$~mJy~\kms. Given that the flux density of the continuum at $230$~GHz is $304$~\mJy, the line has peak optical depth $\tau_{\rm peak}=0.003$ and integrated optical depth $\int\tau dv =0.221$\kms. Our observations spectrally resolve the absorption line and we measure a Full Width at Half Maximum, FWHM$=57.4\pm20$\kms.

In the centre, if the \co\ was regularly rotating within the disk, we would have detected it at the systemic velocity, given the small size of the radio source. Instead, the absorption line peaks at $v_\mathrm{peak} = 4615\pm20$\kms and has an offset with respect to the systemic velocity of $\Delta v = v_\mathrm{peak} - v_\mathrm{sys} = + 365\pm22$\kms (see Table~\ref{table:abs}). Since the gas is detected in absorption it is located in front of the radio source, it is falling towards it and the in-fall velocity is given by the shift of the line $v_\mathrm{infall} \lesssim  \Delta v = + 365\pm22$\kms.

If we assume that the gas detected in absorption is homogeneously distributed in front of the background continuum source, and that it is in local thermal equilibrium (LTE), we can determine the column density of the carbon monoxide in the level $J=2$   ($N_{\rm CO\, (2)}$) from the integrated optical depth of the absorption line ($\int \tau dv$) using equation (\eg\ \citealt{wiklind1995,bolatto2013}): 

\begin{equation}
N_\mathrm{CO\, (J)} =  \frac{3h}{8\pi^3\mu^2}\frac{g_J}{J\cdot\bigg(e^{\frac{h\nu_{\rm rest}}{k_BT_\mathrm{ex}}}-1\bigg)} \cdot \int \tau_\mathrm{CO\,(J)} dv
\end{equation}

where $\mu=0.112$~cm$^{5/2}$~g$^{1/2}$~s$^{-1}$ is the dipole moment of the CO molecule, $J=2$ is the upper level of the $J\rightarrow J-1$ transition, $g_J=2J+1$ is the statistical weight of the transition, $T_\mathrm{ex}$ its excitation temperature and $\nu_{\rm rest}$ its rest frequency, $h$ and $k_B$ are the Planck's and Boltzmann constants, respectively. 
To determine the column density of the carbon monoxide we need to assume its excitation temperature. 

In the Milky Way, in conditions of thermal equilibrium the typical temperature of the CO gas is $\sim 16$ K (\eg\ \citealt{heyer2009}). Nevertheless, close to the nuclear activity, the gas could be warmer, which would increase the column density. For example, in the circumnuclear regions of \mbox{4C +12.50} the gas has likely temperature $T_\mathrm{ex}\sim60$ K~\citep{dasyra2012,dasyra2014} which would increase the column density of the \co. Here, for simplicity, we assume the \co\ is in thermal equilibrium at $T_\mathrm{ex}\sim16$ and the column densities we derive are lower limits, if the temperature of the gas was higher. 

\begin{table}
	\caption{Properties of the \co\ detected in absorption}             
	\label{table:abs}      
	\centering                          
	\begin{tabularx}{\columnwidth}{X c }        
		\hline\hline                 
		Parameter 						 & Value									\\
		\hline
		S$_{\rm 230\, GHz}$ 			 & $304\pm30.4$\mJy	 								\\
		S$_{\rm abs}$ 					 & $2.6\pm 0.3$ \mJy 								\\
		S$_{\rm peak}$					 & $-0.8\pm0.1$ \mJy								\\
		FWHM 							 & $57\pm20$\kms 									\\
		$\Delta v = v_{\rm peak} -$\vsys & $345\pm20$\kms									\\
		Peak optical depth 			  	 & 0.003									\\
		Integrated optical depth		 & 0.221 \kms 								\\
		$N_{\rm CO}$			 		 & $ \gtrsim 1.0\times10^{17}$~\cmsq \\
		$N_{\rm H_2}$			 		 & $\gtrsim 1.6\times10^{21}$\cmsq	\\
		\mhtwo\	($r$ = 2 --75 pc)& $3\times10^2$~\msun\  -- $5\times10^5$~\msun \\
		\hline
	\end{tabularx}
\end{table}

The column density of the \co\ gas is $N_\mathrm{CO\, (2)} = 9\times10^{14}$~\cmsq, respectively. To derive the column density of the total CO gas, we multiply $N_\mathrm{CO\, (2)}$ by the partition function of the 2-1 rotational transition of the CO gas, $\sum_J (2\mathrm{J}+1) e^{-E_\mathrm{J}/k_B\,T} = \frac{k_B\,T_\mathrm{ex}}{B}$ (where $B$ is the the rotational constant of the molecule expressed in cm$^{-1}$). We divide the result by the relative population of the upper level of the transition, $g_J\, \mathrm{exp}(-E_u/k_B\,T_\mathrm{ex})$ (where $E_u=11.5$ cm$^{-1}$), obtaining $N_\mathrm{CO} = 1.0\times10^{17}$~\cmsq. To convert it to the column density of the molecular hydrogen we assume a CO-to-\Htwo\ abundance ratio, $A_C= 1.6\times10^{-4}$~\citep{sofia2004}, and we find $N_\mathrm{H_2}= 1.6\times10^{21}$~\cmsq. This result is compatible with the $3\sigma$ detection limit of the \co\ line emission on the same line of sight ($N_{\mathrm H_2}\sim4.8\times10^{21}$, see Sect.~\ref{sec:indisk}).

The width of the absorption line also suggests that the absorbing \co\ is part of a large and complex ensemble of molecular clouds. In the spectrum at the original channel width of the observations ($10.3$ \kms), we resolve the absorption line over five channels. The width of the absorption line is much higher than the typical velocity dispersion of the molecular gas in a cloud ($\lesssim 4 $\kms,~\eg\ \citealt{heyer2009}). According to the scaling relations for molecular clouds in the normal ISM~\citep{larson1981,solomon1987,heyer2009}, the size of a cloud corresponding to the width of the absorption line would be $\sim 1$ kpc, which is the extent of the entire circumnuclear disk. Hence, the absorption line is likely tracing an ensemble of transient clouds, that are not in equilibrium with the surrounding medium, have non-circular kinematics, and are being shredded and torn apart while moving towards the SMBH (see Sect.~\ref{sec:infall} for further details). 

We estimate the range of possible masses of the absorbed in-falling molecular gas assuming that the gas could extend only exactly over the background radio continuum emission ($r\sim2$ pc) or that it could fill the entire region where \co\ emission is not detected ($r\sim 75$ pc). Given these two extreme possible cases, for a column density of $\sim 1.6\times10^{21}$\cmsq (see Table~\ref{table:abs}), the mass of the in-falling molecular gas ranges between $3\times10^2$~\msun\ and $5\times10^5$~\msun.

\subsection{Luminosities and masses of the molecular gas}
\label{sec:masses}

The emission of the \co\ line allows us to determine the molecular masses of the different structures we identified in the previous sections,~\ie\ the mass derived from the total flux density of the \co\ line in \pks\ and from the flux of the circumnuclear disk (see Table~\ref{table:em}). 

To convert the flux of the \co\ line to mass we need to assume the ratio between the luminosity of the observed line and the \couno\ line. Optically thick gas in thermal equilibrium has brightness temperature and line luminosity independent of the rotational energy level $J$ of the line, $L'_{\rm CO (2-1)}=L'_{\rm CO (1-0)}$ (in units of K \kms pc$^2$). If the gas was optically thin, the ratio could be higher than one. Here, we are not able to set constraints on the optical thickness of the \co\ (observations of different excitation levels, or of different isotopologues of the CO, would be needed) and, for simplicity, we assume $L'_{\rm CO (2-1)}=L'_{\rm CO (1-0)}$. If instead the molecular gas was sub-thermally excited, the values we derive for the luminosity of the total $^{12}$CO would be lower limits. 

According to \citealt{solomon2005,bolatto2013} (and references therein), the luminosity of the CO line is given by:

\begin{align*}
L'_{\rm CO}\,\, [{\rm K\, km\,s}^{-1}\, {\rm pc}^2] = &\,\, 3.25\times10^{7}\,S_{\rm CO}\,\Delta v\, [{\rm Jy\,km\,s}^{-1}]\,\times \\
													 &\,\,\times\nu_{\rm obs}^{-2}\,[{\rm GHz}] \times \frac{D_L^2\,[{\rm Mpc}]}{(1+z)^3}
\end{align*}

\noindent where $\nu_{\rm obs}$ is the frequency at which the line is observed, and $D_L$ is the luminosity distance of the source in Mpc. The total luminosity of the \co\ emission in \pks\ is $L'_{\rm CO}=7.4 \times 10^7$~K~km~s$^{-1}$~pc$^2$ and the circumnuclear disk has luminosity of $L'_{\rm CO}= 1.2\times 10^7$~K~km~s$^{-1}$~pc$^2$.

From these we determine the total mass of the molecular hydrogen traced by the \co\ line and the molecular hydrogen mass of the circumnuclear disk. The \Htwo\ mass-to-CO luminosity relation can be expressed as $M_{{\rm H}_2} =$ \alphac$L'_{\rm CO}$ \citep{solomon1997,downes1998,solomon2005,dale2005,bolatto2013}. In the Milky Way, most of the CO is detected in optically thick molecular clouds and \alphac$= 4.6$~\alun. If the temperature of the gas was higher, \eg\ in Ultra Luminous InfraRed Galaxies (ULIRGs) and AGN, the conversion factor could be as low as \alphac$=0.8$~\citep{downes1998,bolatto2013,geach2014}. In Table~\ref{table:em}, we show that the total mass of the \Htwo\ varies between $3.26\times10^8$~\msun\ for (\alphac$=0.8$~\alun), to $1.88\times10^9$~\msun\ for a Galactic conversion factor (\alphac$=4.6$~\alun). The mass of the circumnuclear disk ranges between $5.35\times10^7$~\msun\ and $3.08\times10^8$~\msun. 

\begin{table*}[tbh]
	\caption{Properties of the \co\ detected in emission}             
	\label{table:em}      
	\centering                          
	\begin{tabular}{l c c }        
		\hline\hline                 
		Parameter 				&											& Value									\\
		\hline                        
		\fluxco\				&	total 								& $174\pm17$ Jy~\kms 						\\
		$L'_{\rm\, CO\,(2-1)}$			&											& $7.3\times 10^7$~K~km~s$^{-1}$~pc$^2$\\
		\mhtwo\ 				&  \alphac $= 0.8$~\alun						& $3.2\times10^8\,\,M_\odot$ 			\\
		&	\alphac $= 4.6$~\alun						& $1.9\times10^9\,\,M_\odot$			\\
		\\
		\fluxco\				&	circumnuclear disk 					& $29\pm3$ Jy~\kms						\\
		$L'_{\rm \,CO\,(2-1)}$			&											& $1.2\times 10^7$~K~km~s$^{-1}$~pc$^2$\\
		\mhtwo\ 				&	\alphac $= 0.8$~\alun						& $5.3\times10^7\,\,M_\odot$			\\
		&   \alphac $= 4.6$~\alun						& $3.1\times10^8\,\,M_\odot$			\\
		\\
	\hline
	\end{tabular}
	\tablefoot{We estimate the total \Htwo\ mass assuming two different conversion factors between the luminosity of the \co\ line and the mass of the molecular gas. The lower values of the mass are obtained adopting the typical value for the optically thick gas in ULIRG and AGN (\alphac$=0.8$~\alun) and the upper limit are given for a Galactic conversion factor (\alphac$=4.6$~\alun). }
\end{table*}


\section{Discussion}
\label{sec:disc_ch6}

\subsection{The \co\ and the \Htwo\ in \pks}
\label{sec:h2}

\begin{figure*}[tbh]
	\begin{center}
		\includegraphics[trim = 0 0 0 0, clip,width=\columnwidth]{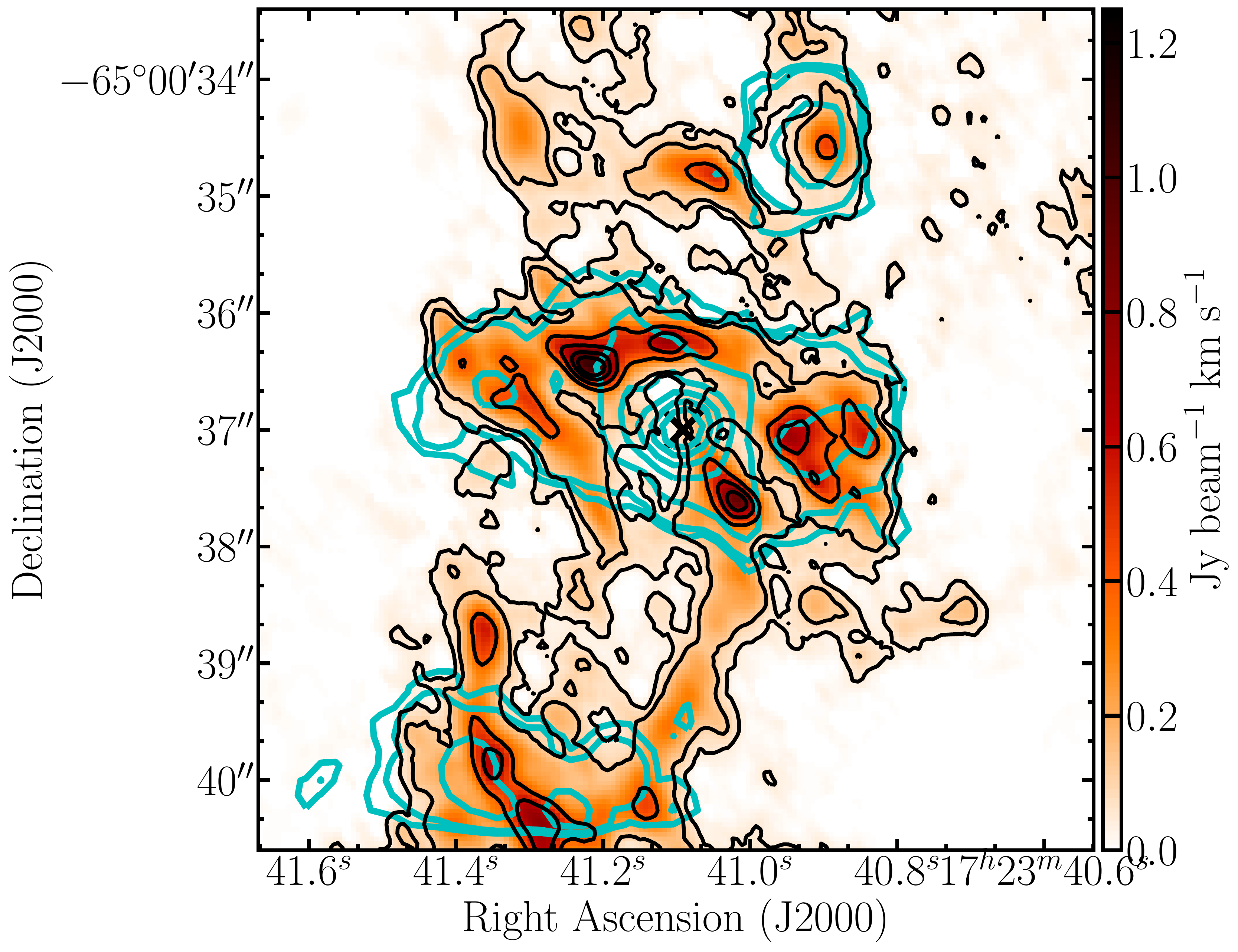}
		\includegraphics[trim = 0 0 0 0, clip,width=\columnwidth]{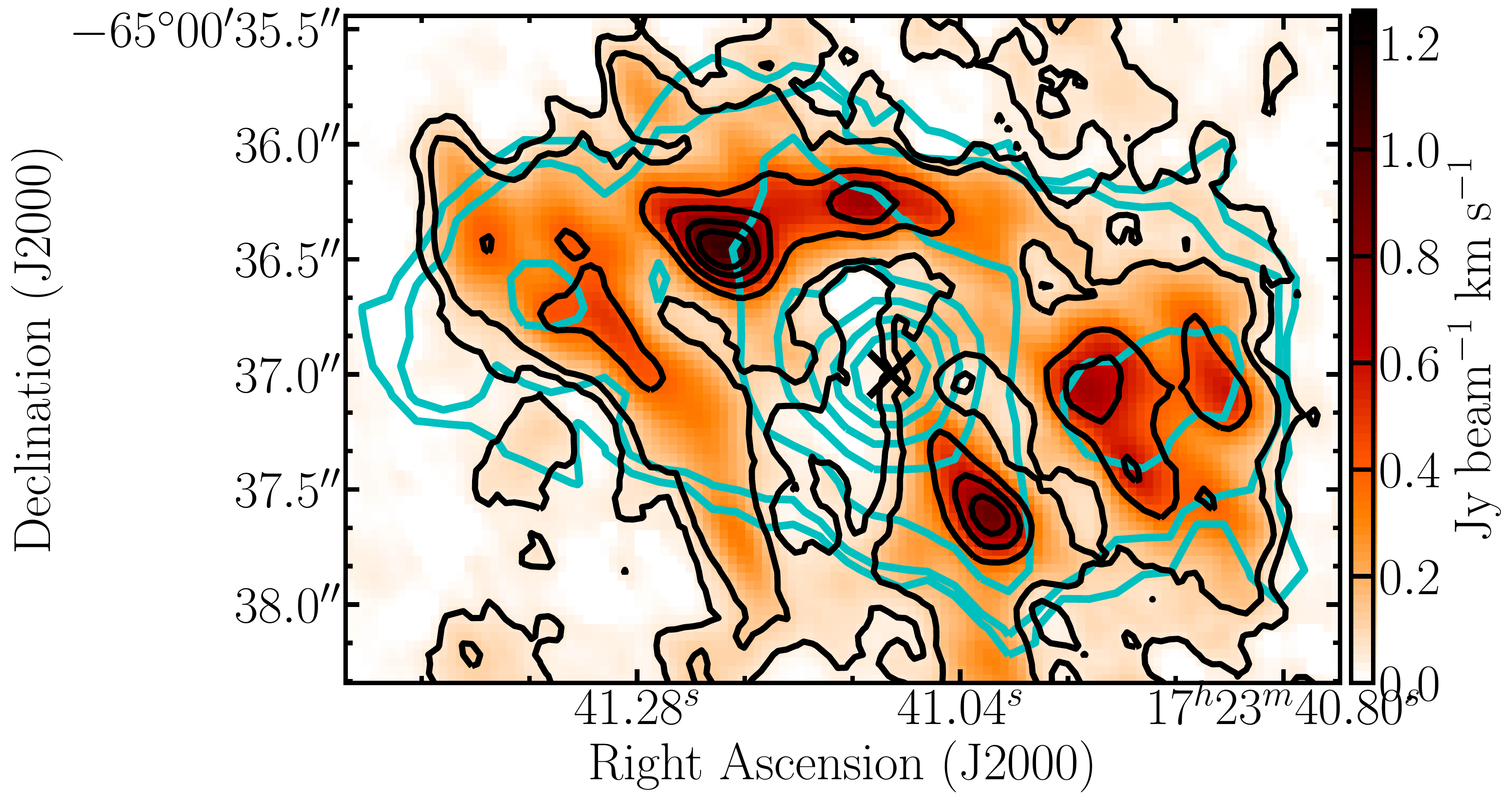}
		\caption{{\em Left panel}: Total intensity image of the distribution of the \co\ gas in \pks\ (black contours) overlaid with the distribution of the \Htwo\ detected by SINFONI, in cyan contours \citep{maccagni2016}. The resolution of the \co\ data is $0.28\arcsec$, the resolution of the \Htwo\ data is $0.55\arcsec$. {\em Right panel}: Total intensity map of the circumnuclear disk of \co\ gas of \pks\ overlaid with the warm \Htwo\ emission. The radio continuum is shown by a black cross.}
		\label{fig:h2_over}
	\end{center}
\end{figure*}

In this section, we compare the SINFONI observations of the warm \Htwo\ \citep{maccagni2016}, with the ALMA observations of the \co. Figure~\ref{fig:h2_over} (left panel) shows the \co\ emission of the circumnuclear disk overlaid with the \Htwoi\ emission detected in the innermost $8$ kpc of \pks. The two lines trace the \Htwo\, in its cold ($T\sim 10-100$ K) and warm ($T\sim 10^3$ K) phase, respectively. The spatial resolution of the SINFONI observations ($0.55\arcsec$) is lower than the ALMA observations. The figure shows that, overall, the warm \Htwo\ and the \co\ gas have a similar spatial distribution within the central region, a circumnuclear disk with radius of $700$ pc oriented east-west and an external disk oriented north-south. The SINFONI observations detected \Htwo\ gas distributed in two orthogonal disks, with possibly gas between the two structures. The ALMA observations trace \co\ gas connecting the two disks, which appear to be part of the same structure.

In the outer disk ($r\gtrsim1$ kpc), the \Htwo\ is detected where the \co\ is brightest. Fig.~\ref{fig:h2_over} (right panel) shows that this occurs also along the circumnuclear disk except for the innermost $75$ pc. There, the \Htwo\ emission is brightest while the \co\ emission is dimmest. In particular, within $75$ pc from the SMBH, the ratio between the column density of the \Htwo\ traced by the \Htwoi\ line and by the \co\ line is ten times higher than in the circumnuclear disk ($N_{\mathrm H_2} \mathrm{(H_2 \, 1-0 \, S(1))} / N_{\mathrm H_2}(\mathrm{ CO \,(2-1)}) = 110$ and $16$, respectively). Since the \co\ and the \Htwoi\ are both tracers of the molecular hydrogen, this suggests that in the centre the conditions of the gas are different than at larger radii. In the centre of \pks, the \co\ line may not be tracing all molecular hydrogen present, but only its coldest component. This can happen when, in the presence of a strong UV-radiative field, but also of X-rays and $\gamma$-rays, the CO gas is photo-dissociated while the \Htwo\ gas is heated (\eg\ \citealt{lamarche2017}). In \pks, a radiation field with this properties could have been recently originated by the newly born nuclear activity and indications of its presence could be found in the excess of radiation in the X-rays and $\gamma$-rays detected in this galaxy~\citep{siemiginowska2016,migliori2016}. If this was the case, we would be observing the radio nuclear activity of a newly born AGN actively changing the physical conditions of the surrounding ISM.

\pks\ is not the only radio AGN where the nuclear radiation may be changing the conditions of the surrounding ISM, and where cold gas clouds of molecular gas may fall onto the SMBH. For example, Centaurus A is a very well studied low efficiency accretion radio AGN where the last episode of nuclear activity was recently triggered ($t_\mathrm{radio}\sim10^6$ yr, see, \eg\ \citealt{morganti2010} for a review). Interestingly, recent ALMA observations of Centaurus A~\citep{espada2017}, reveal a clumpy circumnuclear disk of molecular gas in its innermost $500$ pc. In the centre of the disk, there is a lack of CO gas, while the warm \Htwo\ is detected down to small radii ($r\lesssim 10$ pc). The results we found for \pks\ indicate that this may be due to different conditions of the molecular gas rather than a complete depletion of CO gas in the very central regions. Moreover, the circumnuclear disk of Centaurus A may be strongly warped, as the one we detect in \pks. In the innermost regions of the disk ($r\sim 10-50$ pc) different tracers (CO, HCN, HCO$^+$ (4--3)) reveal cold clouds of molecular gas with non circular orbits.

\begin{figure}[tbh]
	\begin{center}
		\includegraphics[width=\columnwidth]{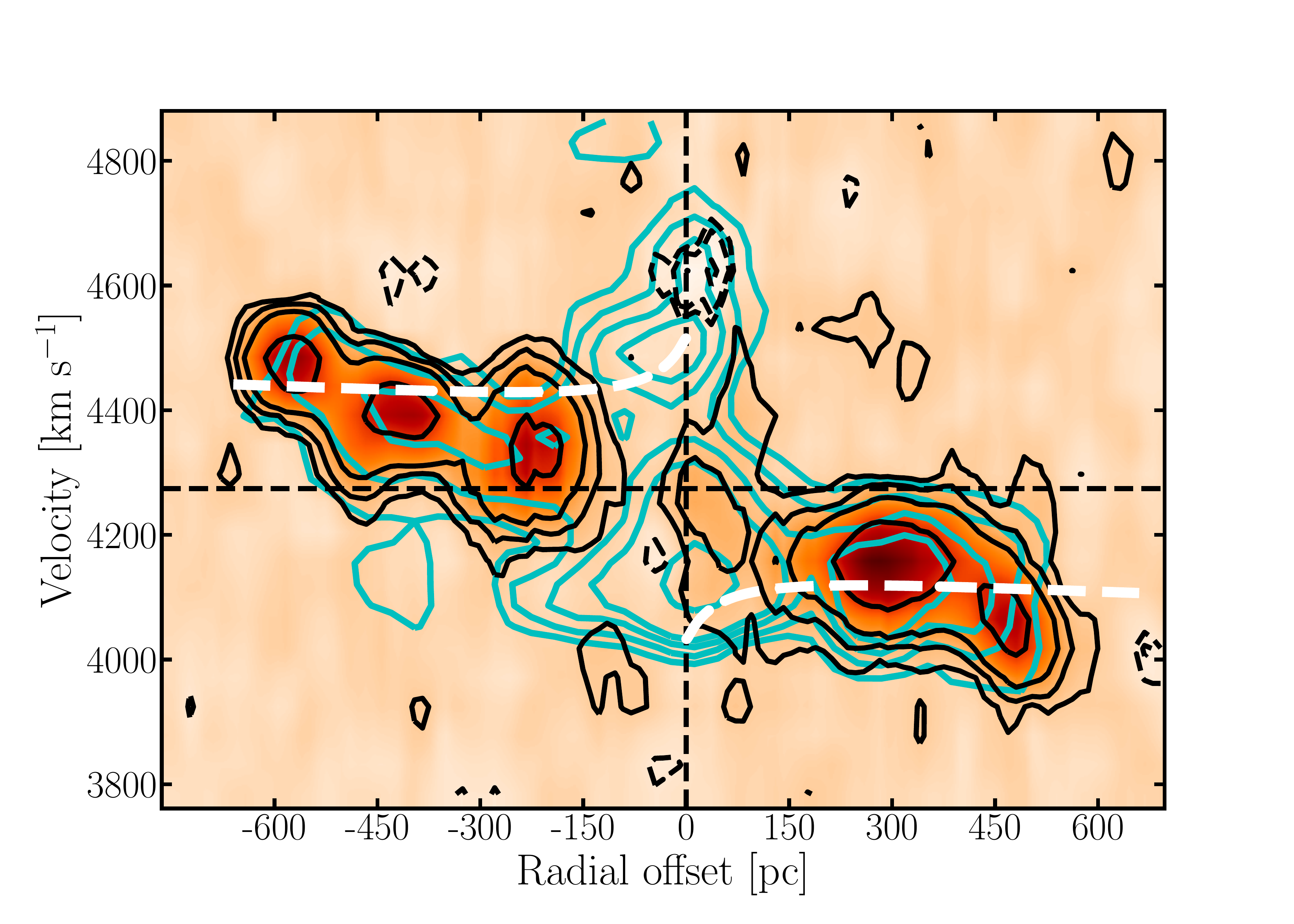}
		\caption{Integrated position-velocity diagram taken along the major axis of the circumnuclear disk \pks\ (PA $=72^\circ$), integrated over  $0.5''$ in the direction perpendicular to the major axis. Black solid and dashed contours show the \co\ emission and absorption, respectively. Cyan contours show the \Htwo\ emission. Contour levels are $-5,-3,\,\,-2\sigma$ and $3,6,12,18,24\sigma$, respectively. The white dashed line shows the rotation curve predicted from the stellar distribution and the contribution of the SMBH to the rotation (see the text for further details).}
		\label{fig:pvplot}
	\end{center}
\end{figure}

Figure~\ref{fig:pvplot} shows the position-velocity diagram extracted along the major axis of the circumnuclear disk and integrated over a width of $0.5\arcsec$ in the direction perpendicular to the major axis. In \citet{maccagni2016}, we estimated a model of the rotation curve of the inner disk (shown by the white dashed line in Fig.~\ref{fig:pvplot}) considering the contribution of the central SMBH and of the stellar mass distribution ($M_\star =4\times10^{11}$\msun, effective radius $r_e= 9.7$ kpc). The rotation curve matches the smooth velocity gradient of the gas, hence, overall, the warm \Htwo\ and the \co\ gas regularly rotate in the circumnuclear disk. Nevertheless, in the centre, there is gas that is deviating from the rotation at red-shifted velocities, as also seen in the SINFONI data \citep{maccagni2016}. 

Overall, the warm \Htwo\ has kinematics characterised by regular rotation, except that in the central resolution element where warm \Htwo\ emission is detected at the same red-shifted velocities of the \co\ gas detected in absorption ($\Delta v = +365\pm20$~\kms, see Sect.~\ref{sec:red_abs}). This is the only component of \Htwo\ gas in the entire field of view with kinematics deviating from rotation, and it appears to be kinematically connected to the circumnuclear disk (see Fig.~\ref{fig:pvplot}). This likely suggests that the \Htwo\ gas with red-shifted velocities lies within the innermost $75$ pc of the galaxy. Given that the \co\ absorption is found on the same line of sight of this \Htwo\ gas and that the two components have the same red-shifted velocities, they are likely located in the same region. Hence, the upper limit on the projected-distance from the SMBH of the in-falling \co\ gas is $d_\mathrm{infall}\lesssim75\pm30$ pc.

\subsection{Molecular clouds fuelling the newly born radio source}
\label{sec:infall}

\begin{figure}[tbh]
	\begin{center}
		\includegraphics[trim = 0 0 0 0, clip,width=\columnwidth]{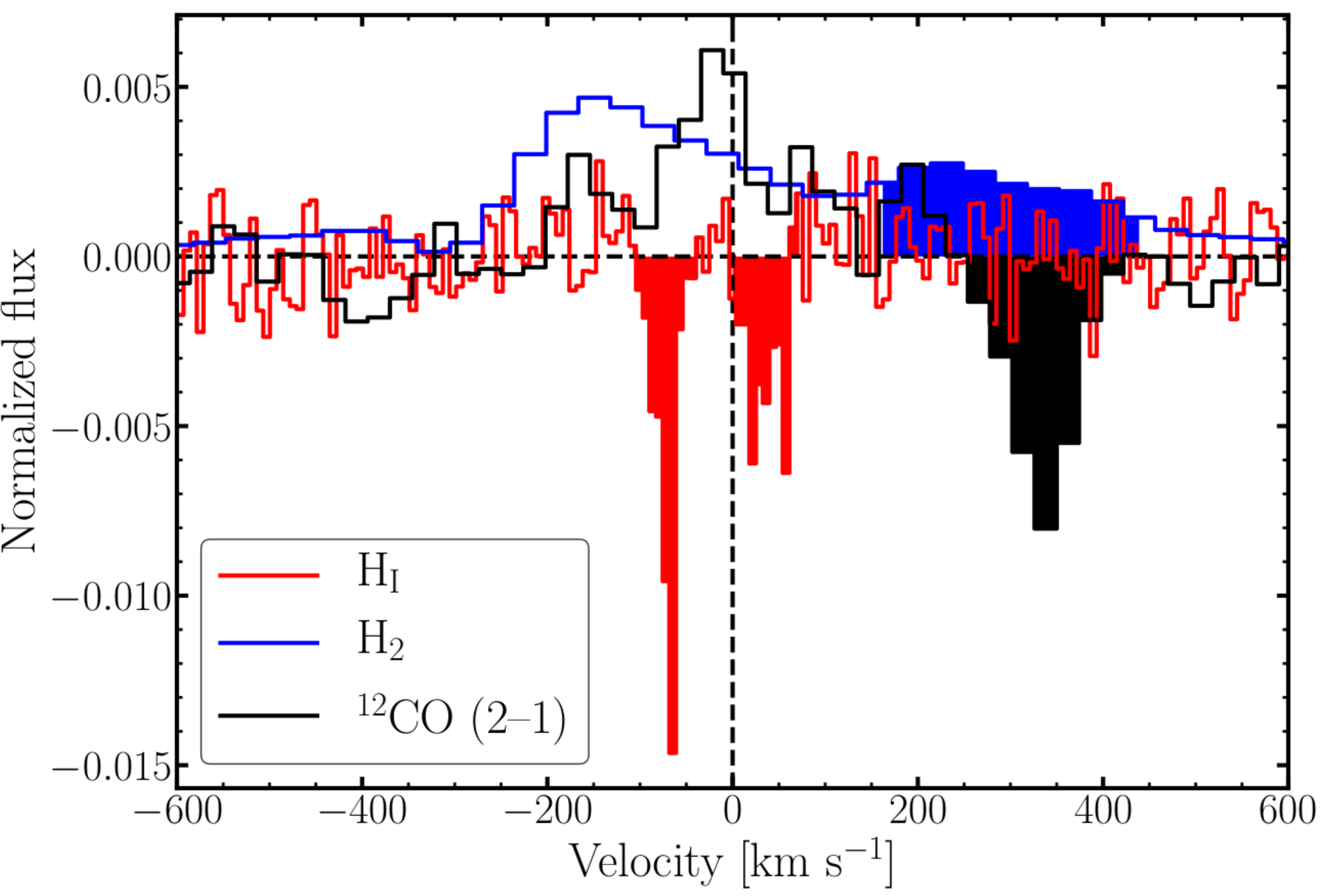}
		\caption{Spectra of the \HI\ (red), \Htwo\ (blue) and \co\ (black) gas extracted against the compact radio continuum emission of \pks. Two separate lines are detected in the \HI\ gas with opposite velocities with respect to the systemic. The \co\ detected in absorption peaks at red-shifted velocities ($\Delta v\sim 343$\kms) with respect to the systemic (black dashed line). The \Htwo\ gas is detected in emission at the systemic velocity and at the velocities of the red-shifted absorbion line of \co\ gas. The spectrum of the \Htwo\ gas has been normalized to the maximum of the spectrum of the \co\ line (see Sect.~\ref{sec:red_abs} for further details). The shaded regions of the lines are tracing gas that is not regularly rotating.}
		\label{fig:spec_overlay}
	\end{center}
\end{figure}

Figure~\ref{fig:spec_overlay} shows the spectrum of the \co\ extracted against the central compact radio source ($r_{\rm radio} \sim 2$ pc, Sect.~\ref{sec:red_abs}) overlaid with the spectrum of the \HI\ \citep{maccagni2014} and of the warm \Htwo\ gas \citep{maccagni2016} extracted along the same line of sight. The spectrum of the \Htwoi\ gas is normalised to the peak of the \co\ emission line. The shaded regions of the figure highlight the parts of the lines tracing gas that is not regularly rotating within the circumnuclear disk. The two \HI\ absorption lines cannot rotate with the disk because they trace gas that is on a line of sight of only $2$ pc and has opposite velocities with respect to the systemic velocity ($v_\mathrm{peak, blue} = -74$\kms\ and $v_\mathrm{peak, red}= +26$\kms, respectively). The warm \Htwo\ gas has a component of the line with red-shifted velocities with respect to the systemic velocity ($v_\mathrm{peak, H_2}= +4600$\kms) and it extends at the same velocities of the \co\ absorption line outside of the range of rotational velocities of the circumnuclear disk (see Sect.~\ref{sec:h2}). The spectra show a first order similarity, despite the differences in velocity and spatial resolution (\ie\ $6.7$~\kms and $\sim11\arcsec$ for the \HI, $75$\kms\ and $\sim0.55\arcsec$ for the \Htwo\ and $24$\kms and $\sim0.28\arcsec$ for the \co).

Figure~\ref{fig:spec_overlay} shows that clouds of cold gas are found in various tracers (\HI, \Htwo\ and \co) at different velocities with respect to the systemic velocity, along the same (very narrow) line of sight of the radio continuum. The clouds are embedded in a circumnuclear disk, and have kinematics with clear deviations from the regular rotation of the disk. This suggests that in the nuclear region of \pks\ there is a reservoir of clouds of cold gas with strong radial motions. Some of these clouds have red-shifted velocities and are located in front of the SMBH, suggesting they are likely falling onto it.

In Sect.~\ref{sec:masses}, we provide a range of the mass of the absorbing molecular clouds, $3\times10^2$~\msun\ -- $5\times10^5$~\msun. Given that the in-fall velocity is $v_{\rm infall}=+365\pm20$~\kms\ and assuming that all the absorbing \co\ gas is at the maximum distance of in-fall, $d_\mathrm{infall}\sim 75$ pc (see Sect.~\ref{sec:h2}), the time-scale of accretion is $t_\mathrm{accretion}\sim5\times10^5$ years. The accretion rate onto the SMBH is $1.6\times10^{-3}$~\msunyr\ $\lesssim\dot{M}_{\rm H_2}\lesssim2.2$~\msunyr. This range is broad because we consider two extreme approximations for the extent and column density of the in-falling clouds (see Sect.~\ref{sec:red_abs}).

\subsection{Accretion of molecular gas in radio AGN}
\label{sec:accretion}

In the previous sections, we defined the main properties of the in-falling clouds of molecular hydrogen that we have detected in the innermost $75$ pc of \pks. Here, we relate these properties to the accretion mechanism that may have triggered the radio source and may still fuel it. In low-efficiency radio AGN, the energetic output of the nuclear activity provides an estimate of the accretion rate required to sustain the AGN. The luminosity at $1.4$ GHz allows us to measure the mechanical energy released by the radio jets (\eg\ \citealt{cavagnolo2010}), while the luminosity of the the \OIII\ line provides information on the radiative energy of the AGN (\eg\ \citealt{allen2006,balmaverde2008,best2012}). 

In \pks, the expected accretion rate is $\dot{M}\sim 10^{-2}$~\msunyr \citep[see][]{maccagni2014}. The range of accretion rates we determine $1.6\times10^{-3}$~\msunyr\ $\lesssim\dot{M}_{\rm H_2}\lesssim2.2$~\msunyr\ is compatible with the expected accretion rate of \pks. Likely, the accretion of all atomic and molecular gas with strong radial motions in the innermost $75$ pc of \pks\ (\ie\ \HI\ clouds detected in absorption, and molecular clouds traced by the \Htwoi\ and \co\ with red-shifted velocities) can fuel the nuclear activity. Even though the difference between the upper limit on the accretion of the \co\ clouds and the expected accretion rate suggest that not all gas within the $75$ pc is directly accreting onto the SMBH.

Chaotic cold accretion could be the triggering and fuelling mechanism of \pks. Different numerical simulations of chaotic cold accretion~\citep{king2015b,gaspari2015b,gaspari2017a,gaspari2017} show that small clouds of multi-phase gas condense out of the hot gas turbulent eddies of the halo of the host galaxy. Overall, this gas would not show kinematics with clear deviations from circular orbits at kiloparsec distance (as found for example in \mbox{PKS 2322-12},~\citealt{tremblay2016}; \mbox{NGC 5044}, \citealt{david2014}; \mbox{PKS 0745-191}, \citealt{russell2016} and Centaurus A, \citealt{espada2017}), while within a few Bondi radii, \ie\ the radius of influence of the SMBH, the inelastic collisions between the clouds are so intense to cancel angular momentum so that accretion onto the SMBH may occur. This breaks the spherical symmetry assumed so far in models of low-efficiency accretion, and can boost the efficiency of accretion in~radio~AGN. Chaotic cold accretion is predicted to manifest through giant molecular associations of clouds (with size between $50$ and $100$ pc and mass up to several $10^5$\msun), with strong non circular orbits. \citet{gaspari2017,gaspari2016} predict that structures falling into radio AGN should appear as \HI\ or CO absorption features. 

The in-falling clouds we detect in the centre of \pks, as well as the different physical conditions of the \HI, \Htwoi\ and \co\ gas in the innermost $75$ pc compared to the circumnuclear disk and the complex warped disk to which it is connected, all well match the properties of the accreting gas predicted by the models of chaotic cold accretion. Moreover, the maximum distance from the AGN where the in-falling gas could be located is $75$ pc. This is less than twice the radius within which the SMBH dominates the kinematics of the galaxy ($r\sim 45$ pc, assuming that the velocity dispersion of the stars is $\sim 200$\kms and $M_\mathrm{SMBH}\sim 4\times10^8$\msun, \citealt{willett2010}). This is also in good agreement with the predictions of chaotic cold accretion models (\eg\ \citealt{gaspari2017a,gaspari2017}). Therefore, it is possible that chaotic cold accretion of molecular clouds triggered the radio nuclear activity of \pks, and is currently fuelling it.

\pks\ could be an optically elusive AGN~\citep{schawinski2015}, \ie\ the AGN is weak in the optical band (classified as LINER) because it is living an intermediate phase of nuclear activity, where accretion has already triggered the mechanism of radiation of radio and X-ray emission, but did not have enough time to sufficiently ionize the circumnuclear ISM to be bright in the optical. Given the age of the radio source, the radiation field of the nuclear activity could have influenced the ISM only within the innermost $\sim28$ pc. Even though the information on the ionized gas surrounding the AGN is limited (\eg\ \citealt{filippenko1985}), \pks\ is optically a LINER AGN, it has a thermal X-ray excess \citep{siemiginowska2016}, and has a young powerful radio AGN carving its way through a dense gas, which is mostly molecular.

\section{Conclusions}
\label{sec:conc_ch6}

In this paper, we presented new ALMA observations of the \co\ gas in the central $15$ kpc around the young radio source \pks. These observations follow the \HI\ \citep{maccagni2014} and warm \Htwo\ \citep{maccagni2016} observations that revealed cold gas in the centre of the galaxy that may contribute fuelling the radio AGN. The ALMA observations allow us to trace with high spatial resolution ($0.28\arcsec$) the distribution of the carbon monoxide, study its kinematics in relation to the recent triggering of the radio nuclear activity, and detect in-falling molecular clouds that may accrete onto the SMBH.

The ALMA observations of the \co\ line reveal that multiple clouds are distributed in a complex warped disk. Between $7.5$ kpc and $2.5$ kpc from the central radio AGN, the disk is oriented as the dust lane of the galaxy in the north-south direction ($PA\sim180^\circ$). At inner radii ($2.5$ kpc - $700$ pc), the disk changes slightly in orientation ($PA\sim150^\circ$) and molecular clouds are found in proximity of dust (Fig.~\ref{fig:mom0}) and of warm \Htwo\ gas (see the left panel of Fig.~\ref{fig:h2_over}). In the very central regions ($r\sim 700$ pc), the disk changes abruptly its orientation and the molecular clouds form an edge on circumnuclear disk oriented approximately east-west ($PA\sim72^\circ$).

The carbon monoxide is a tracer of the cold phase of the molecular hydrogen. Depending on the conversion factor assumed, the total mass of cold molecular hydrogen is \mhtwo\ $3.2\times10^8$--$1.9\times10^9$~\msun\ (see Table~\ref{table:em}), while the circumnuclear disk has total mass of cold molecular hydrogen between $5.3\times10^7$~\msun\ and $3.1\times10^8$~\msun.

The circumnuclear disk of molecular gas of \pks\ may form the reservoir of fuel for the radio emission. Although the rotation of the gas follows what expected from the mass of the central SMBH and the distribution of the stellar body, the velocity dispersion of the gas is higher in the centre of the disk ($r<150$ pc, $0.45\arcsec$) than at larger radii (Fig.~\ref{fig:pv_spec}). This suggests that close to the radio source the gas may have non circular orbits. 

The circumnuclear disk is also traced by the \Htwoi\ line emission \citep{maccagni2016}. The ratio between the column density of the \Htwo\ gas traced by the \Htwoi\ line and by the \co\ line increases by a factor ten in the innermost $75$ pc of the circumnuclear disk with respect to its outer regions ($\sim110$ and $\sim16$ respectively). This suggests that, in the centre, the physical conditions of the molecular gas are different. It is possible that the radiative field of the AGN is photo-dissociating the cold component of the molecular gas, traced by the \co\ line, before the warmer component, traced by the \Htwoi\ line. This would indicate that the newly born AGN is actively changing the physical conditions of the surrounding ISM.

Against the $230$ GHz emission of the radio AGN, we detect \co\ gas in absorption (see Sect.~\ref{sec:red_abs}). The absorption line has a width (FWHM$=54$\kms) much larger than the typical velocity dispersion of molecular clouds. Hence, either the absorption is tracing multiple clouds, or a single cloud with non circular orbits, or a combination of both. The absorption line is detected at red-shifted velocities with respect to the systemic velocity ($\Delta v = +365\pm20$\kms). Its kinematics deviate from the rotation of the circumnuclear disk and the gas may be falling towards the radio source. At the same red-shifted velocities we also detect in emission a component of warm \Htwo\ (see Fig.~\ref{fig:pvplot}). This allows us to constrain the distance from the SMBH of the red-shifted \co\ to $75$ pc. Likely, these clouds are accreting onto the SMBH. This is one of the best indications that accretion of cold molecular clouds can trigger and fuel a young radio source. These clouds may have originated from condensation of the hot gas present in the centre of \pks, as suggested by the excess in soft X-ray emission \citep{siemiginowska2016}. In Section~\ref{sec:infall}, we suggest that the in-falling molecular clouds may trace chaotic cold accretion~\citep{gaspari2017a,gaspari2017,gaspari2016} onto the SMBH of \pks. 

\pks\ is the youngest radio source showing strong indications that cold gaseous clouds are accreting onto the SMBH. Indications that chaotic cold accretion can fuel an AGN have been found also in a handful of other sources. For example, in \mbox{Centaurus A} \citep{espada2017}, \mbox{NGC 5044} \citep{david2014,russell2016,david2017} and \mbox{PKS 2322-12}~\citep{tremblay2016} cold gas clouds with non circular orbits are found in proximity of the AGN. While in \mbox{NGC 3115} \citep{wong2013} and \mbox{M 87} \citep{russell2015} similar clouds are likely condensating from the hot circumnuclear environment.

The ALMA observations of only the \co\ transition do not allow us to constrain the physical conditions of the circumnuclear disk, such as temperature, pressure and heating mechanism of the gas. This would be a crucial step forward in the study of accretion of cold gas in low-efficiency radio AGN, and in understanding how the radio nuclear activity can change the conditions of the surrounding molecular gas. For these reasons, we planned to continue the study of the molecular gas in \pks\ and its contribution to fuel the newly born radio AGN with follow-up ALMA observations of the \cotre\ and HCO+ (4--3) at the same spatial and spectral resolution. We will determine where along the disk the molecules are heated by photo-dissociation and where by X-ray dissociation, and we will measure if there are gradients in pressure between the clouds of the circumnuclear disk. This will allow us to relate the properties of the molecular gas to the radio and X-ray properties of the AGN and, possibly, to set further constraints on the processes of accretion onto the SMBH of young radio sources.


\begin{acknowledgements}
ALMA is a partnership of ESO (representing its member states), NSF (USA) and NINS (Japan), together with NRC (Canada), NSC and ASIAA (Taiwan), and KASI (Republic of Korea), in cooperation with the Republic of Chile. The Joint ALMA Observatory is operated by ESO, AUI/NRAO and NAOJ. The National Radio Astronomy Observatory (NRAO), which is a facility of the National Science Foundation operated under cooperative agreement by Associated Universities, Inc.The research leading to these results has received funding from the European Research Council under the European Union's Seventh Framework Programme (FP/2007-2013) / ERC Advanced Grant RADIOLIFE-320745. The authors wish to thank Dr. Massimo Gaspari for the useful discussions and suggestions.
\end{acknowledgements}



\bibliographystyle{aa}
\bibliography{phd_biblio3}

\end{document}